\documentclass[acmtomccap]{acmlarge}\acmNumber{0}
\acmArticle{0}
\articleSeq{0}
\acmYear{0}
\acmMonth{0}
\usepackage[ruled]{algorithm2e}
\usepackage{amsmath}
\usepackage{subfigure}
\usepackage{caption}
\usepackage{multirow}
\usepackage{listings}
\SetAlFnt{\algofont}
\SetAlCapFnt{\algofont}
\SetAlCapNameFnt{\algofont}
\SetAlCapHSkip{0pt}
\IncMargin{-\parindent}

\renewcommand{\baselinestretch}{0.95}

\title{Saving Energy in Mobile Devices for On-Demand Multimedia Streaming  -- A Cross-Layer Approach}

\author{Mohammad Asharful Hoque, Matti Siekkinen, Jukka K. Nurminen, Sasu Tarkoma, and Mika Aalto}
\begin{abstract}
This paper proposes a novel energy-efficient multimedia delivery system called EStreamer. First, we study the relationship between buffer size at the client, burst-shaped TCP-based multimedia traffic, and energy consumption of wireless network interfaces in smartphones. Based on the study, we design and implement EStreamer for constant bit rate and rate-adaptive streaming. EStreamer can improve battery lifetime by 3x, 1.5x and 2x while streaming over Wi-Fi, 3G and 4G respectively. 
\end{abstract}



\acmformat{Mohammad A. Hoque, Matti Siekkinen, Jukka K. Nurminen, Sasu Tarkoma, and Mika Aalto. 2013. Saving Energy in Mobile Devices for On-Demand Multimedia Streaming  -- A Cross-Layer Approach.}

\begin{document}

\begin{bottomstuff}
Authors' Affiliation: M. A. Hoque, M. Siekkinen, J.K. Nurminen, Computer Science \& Engineering, Aalto University School of Science, emails: {mohammad.hoque, matti.siekkinen, jukka.k.nurminen}@aalto.fi; M.A.Hoque, S. Tarkoma, Computer Science, University of Helsinki, emails: {mohammad.hoque, sasu.tarkoma}@cs.helsinki.fi; M. Aalto, Nokia Solutions and Networks, email: mika.aalto@nsn.com.
\doi{2556942}
\end{bottomstuff}
\maketitle
\section{Introduction}
\label{sec:introduction}

On-demand multimedia streaming services, such as Spotify, Netflix, Dailymotion, Vimeo and YouTube, have gained great acceptance among smartphone users. It is reported that there are six billion hours of YouTube video views every month and 25\% comes from mobile devices~\cite{youtube:stat}. However, the battery life of smartphones becomes critical when accessing these multimedia services via wireless networks (e.g. Wi-Fi, 3G and 4G). Most often the power consumption of mobile devices for using these wireless networks is equivalent to or higher than the playback power consumption~\cite{hoque2013wowmom}. This is because there is continuous flow of data packets during streaming and a smartphone must keep the wireless network interface (WNI) always active for receiving those packets. In this work, we optimize the wireless communication energy spent by mobile devices for TCP-based multimedia streaming.

It is well known that shaping multimedia traffic into periodic bursts can save energy, specifically for UDP-based multimedia traffic shaping over Wi-Fi~\cite{9.wang}. A number of packets are collected over a period of time and then sent together as one burst to the client using all the available bandwidth. In this way, the WNI is kept active only for a short period of time to receive the burst, instead of keeping the interface always active. When the burst size exceeds the playback buffer, there is packet loss. Therefore, the maximum size of a burst, or equivalently the length of a burst interval, is tuned based on an acceptable range of packet loss~\cite{9.wang}.

Today HTTP over TCP is by far the most prevalent set of protocols used for streaming~\cite{guole}. Traffic shaping saves energy with TCP-based streaming as well~\cite{hoque11ccnc}. TCP-based streaming differs from the case of UDP-based streaming in one major way: TCP is a reliable protocol. When the player buffer becomes full, data accumulates into the TCP receive buffer. As a result, an arbitrarily large burst, which is larger than the playback buffer, can be used without any packet loss. However, the exact impact on the energy consumption of smartphones is not well understood. In this article, we first study the interplay between the burst size and the power consumption of smartphones. Specifically, we model the energy consumption of WNIs for bursty TCP traffic. We show that the power consumption of a smartphone decreases when the receiving burst size increases and as long as the client device can accommodate the entire burst. In contrast, the power consumption rapidly increases, if the burst size is too large compared to the available buffer space at the client.

As a proof of concept, we design and implement an energy-efficient multimedia delivery system called EStreamer. It determines an energy-optimal burst size so that smooth playback of the streaming applications is not distorted. EStreamer relies on standard TCP feedback from the client. We thoroughly evaluate EStreamer through measurements with real streaming services. We focus on two aspects: (a) the potential to save energy in smartphones using Wi-Fi, 3G (HSPA) and 4G (LTE) and (b) the impact on the radio access network (RAN) signaling load when using a cellular access network. 

Concerning the first aspect, we measure the energy savings in four different smartphones while streaming from popular streaming services via EStreamer, such as Internet radio, YouTube, and Dailymotion. The results demonstrate that the largest energy savings can be achieved with Wi-Fi (65\%), followed by 4G (50-60\%), after which comes 3G (38\%). It is also shown that a client can achieve similar range of energy savings when streaming from a rate-adaptive streaming service which is powered by EStreamer.

As for the second aspect of evaluation, the energy savings with traffic shaping arise from the ability of the radio interface to transition to lower power consuming states in between two consecutive bursts of content. The state transitions, specifically when using a cellular access network (HSPA or LTE), require signaling message exchange between the smartphone and the network. The amount of signaling traffic can be a severe problem for the network operators. Globally many major
operators have suffered service quality deterioration and even network
outages because of the signaling storms created by the
smartphones~\cite{signals103g}. Consequently, we measure signaling load in the network when applying traffic shaping with different parameter settings in the network. In summary, this article makes the following contributions:
\begin{enumerate}
\itemsep0em
   \item We uncover the relationship between available buffer space at the client, burst size, and power consumption in TCP-based multimedia traffic shaping. Specifically, we develop novel power consumption models for bursty multimedia traffic over TCP to derive a heuristic for energy-optimal burst size for a given client. We show that our models are independent of the kinds of streaming it is applied for.

   \item We design and implement a cross-layer multimedia delivery system called EStreamer, which implements the heuristic. The heuristic makes EStreamer agnostic to device variation and the wireless access (Wi-Fi, 3G or 4G)  used by a smartphone for multimedia streaming. EStreamer has built-in support also for rate-adaptive streaming in addition to constant quality streaming.

\item We evaluate EStreamer in a number of scenarios focusing on the amount of energy saved and the impact on the RAN
  signaling load. Our study covers four smartphones, both
  audio and video streaming services, and cellular
  networks with different configurations. The results show that
  network parameter values have no impact on the energy
  consumption when streaming audio/video, whereas
  they have a major impact when
  EStreamer is used. We show that there are differences in terms
  of signaling load with optimization mechanisms that provide a similar
  amount of energy savings. For HSPA, the best choice is to use
  shorter inactivity timers with traffic shaping. However, when
  discontinuous reception mechanisms are available, such as DRX in
  LTE, those mechanisms together with long inactivity timers are the
  best choice.

\end{enumerate}

\section{Energy Efficiency of Wireless Multimedia Streaming over TCP}
\label{two}

In this work, we focus on the energy consumption of wireless network interfaces for TCP-based multimedia streaming. We first outline the power consumption characteristics of Wi-Fi and cellular network interfaces in smartphones. Then, we describe the characteristics of mobile multimedia streaming. Finally, we explain how burst-shaped multimedia delivery can save energy.

\subsection{Energy Consumption of Wireless Network Interfaces}
\label{two_one}

Some inactivity timers control the energy consumption of different types of wireless communication. Because of these timers, there is residual time spent by any interface in active state after each transmission or reception for the sake of user experience, which leads to some energy spent doing nothing useful. We refer to this energy as \textit{tail energy}~\cite{balasubramanian09imc3g}. Below, we discuss the power saving mechanisms for the wireless radios in smartphones to access the three most commonly used wireless networks, namely Wi-Fi, HSPA, and LTE.

\subsubsection{Wi-Fi}
802.11 standard includes a Power Saving Mechanism (PSM). Commercial
mobile phones typically implement a slightly modified version of it
which is sometimes referred to as PSM-Adaptive (PSM-A).
This version differs from the original one proposed by the standard in
that it keeps the interface in the \textit{idle} state for a certain
fixed period of time (e.g. 200 ms) instead of switching to the
\textit{sleep} state immediately after the transmission or reception
of traffic~\cite{hoque12survey}. This amount of time also determines the amount of
tail energy in Wi-Fi communication. Usually, the power draw in sleep
state is an order of magnitude lower than in idle state. In both receive and
transmit states, the Wi-Fi interface may consume up to 50\% more power than in the idle state.

\begin{figure}[tp]
\begin{minipage}[b]{0.5\linewidth}
\centering
\includegraphics[width=0.7\linewidth, height=0.45\linewidth]{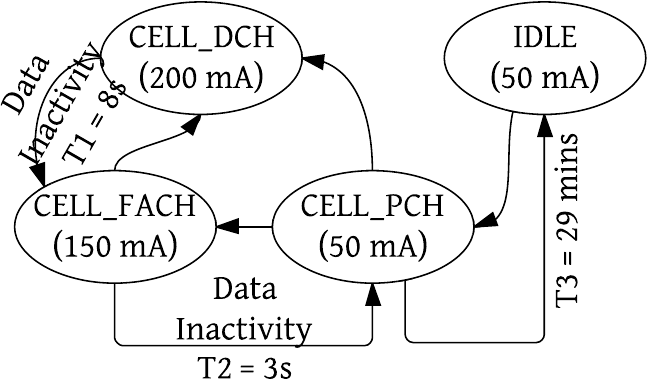}
\caption{3G HSPA state transitions and power consumption.}
\label{fig:3g_stands}
\end{minipage}
\hspace{0.01cm}
\begin{minipage}[b]{0.5\linewidth}
\centering
\includegraphics[width=0.9\linewidth,height=0.45\linewidth]{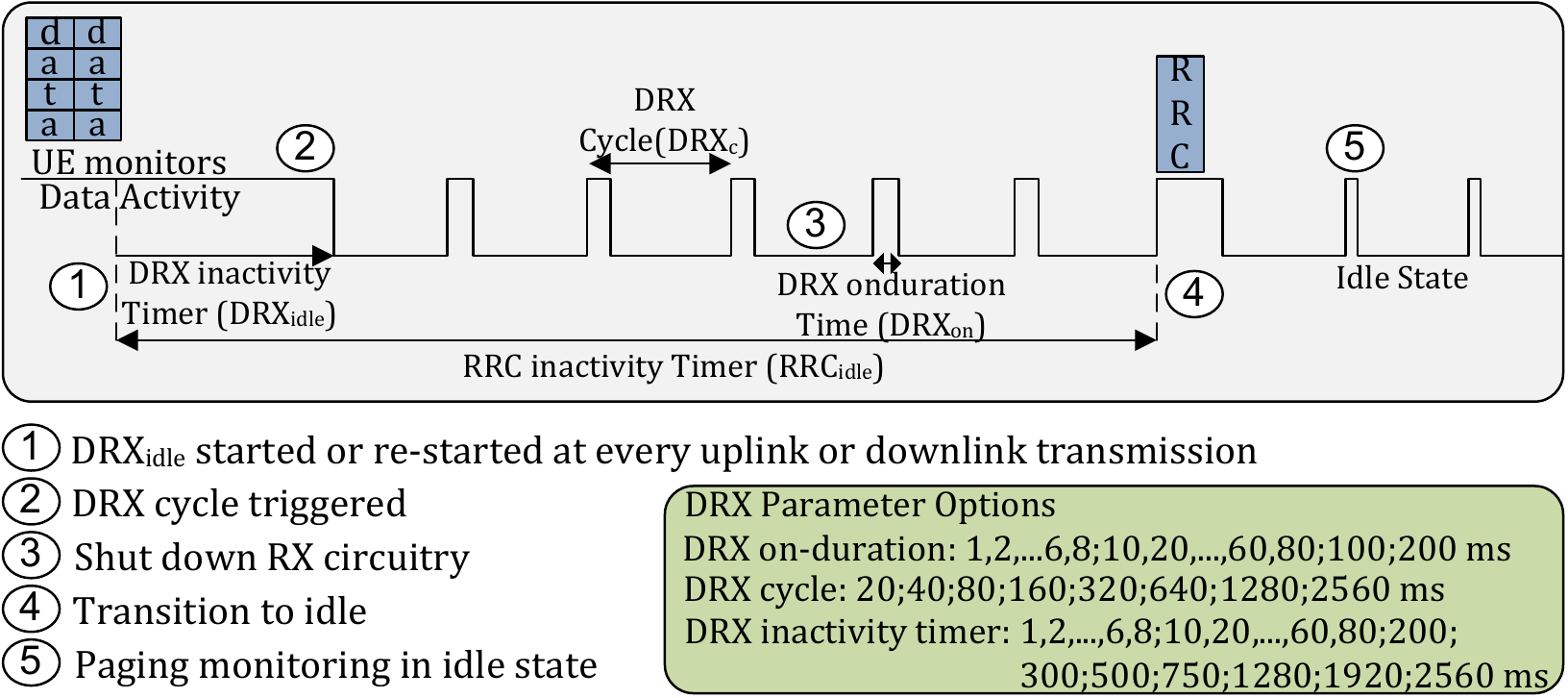}
\caption{LTE RRC with connected state DRX.}
\label{fig:lte_drx_rrc}
\end{minipage}
\vspace{-4mm}
\end{figure}

\subsubsection{HSPA/3G}
HSPA is the widely deployed 3G access technology. In the network, the usage of radio resources and power consumption of a mobile phone is controlled by the Radio Resource Control (RRC) protocol. This protocol has four different states. Figure~\ref{fig:3g_stands} illustrates the RRC state machine and the inactivity timers, which control the transitions among these states. These timer values are in the order of several seconds and controlled by the network operators. A mobile device terminates RRC connection after the CELL\_PCH$\rightarrow$IDLE transition which happens upon expiry of the T3 timer. Some operators may disable CELL\_PCH in their network and in that case the device would terminate the RRC connection when the T2 timer expires. The figure also shows that operating in different states draws different amounts of current. Fast Dormancy (FD) tries to reduce the tail energy effect as the mobile phone can request the network to transition directly CELL\_DCH$\rightarrow$CELL\_PCH (FD Rel-8) or it may terminate the RRC connection and directly end up in IDLE mode (typical for legacy FD).

Different state transitions require some signaling message exchange between a mobile phone and the network. When a device is disconnected from the RRC, the reconnection requires comparatively a lot more signaling than for normal state changes~\cite{nsn2011report}. If the device supports only legacy FD or the network does not support CELL\_PCH then there will be more signaling in the network as they both require RRC reconnection upon data activity.

\subsubsection{LTE/4G}

LTE (Long Term Evolution) is the fastest growing 4G cellular network
technology with commercial networks having already been deployed in
many countries. The LTE RRC protocol contains two states: RRC\_IDLE
and RRC\_CONNECTED. Similar to the HSPA, there is an
inactivity timer with a typical value of 10 s associated to the
transition from the connected to the idle state which may cause a
significant amount of tail energy to be spent. However, LTE includes a
discontinuous reception mechanism specifically to be used in the
RRC\_CONNECTED state, hence called connected state DRX (cDRX) which
can drastically reduce the tail energy. Figure \ref{fig:lte_drx_rrc}
illustrates this mechanism. The idea is that after no packets have
been received for a time period specified by the cDRX inactivity timer
(DRX$_{idle}$), the device starts a duty cycle so that it wakes up
only periodically (every DRX$_c$) for DRX on-duration specified amount
of time (DRX$_{on}$) to check for new incoming packets. If no packets
are received during an interval specified by the inactivity timer
(RRC$_{idle}$), the device transitions from RRC\_CONNECTED to RRC\_IDLE
state. In this way, there are only two states and thus less signaling due to state transitions in the network.

\subsection{Mobile Multimedia Streaming}
\label{sub_character}

In this section, we describe the key properties of on-demand multimedia content and the techniques used by the streaming services to deliver content to the client. This is because some of them have direct impact on the energy consumption of smartphones and thus influence the design of our energy efficient streaming.

\subsubsection{Playback Duration, File Size, Formats and Quality} The key properties of on-demand content are 
the content length, file size, format, and the bit-rate that is an indicator for the playback quality. In the case of audio, the duration of music files are within the range of 3-5 minutes. However, ShoutCast-like services give support for Internet radio stations to provide continuous audio streaming. A number of studies have investigated these properties at the Video-On-Demand (VoD) streaming Web sites, such as YouTube. \citeN{6522525} described that the distribution of video length has three peaks. The first peak is within one minute of playback and consists of more than 20\% of the videos. The second peak is between 3 and 4 minutes and contains about 16.7\% of the videos. The third peak is much smaller and it is near the length of 10 minutes. Thus VoD content typically has a playback duration of several minutes. Recent studies also show that video viewing abandonment affects the video viewing time distribution \cite{Hwang:2012}. This means that the actual viewing time can be less than the video length. Later in Section~\ref{four_two}, we will see how the length of videos influences in selecting a binary search algorithm for energy efficient streaming.

The file sizes were also observed to follow a similar distribution to the video length and be small, typically less than 30 MB. 87.6 \% of the crawled videos used CBR with most videos having a bit-rate around 330 kbps and two other peaks at 285 kpbs and 200 kpbs \cite{6522525}. Many different combinations of the container, encoding algorithm's bit rate, quality, and resolution are in use today. In mobile multimedia services, the most popular combination is H.264-based MP4-360p~\cite{finamore11youtube}, where MP4 is the container and 360p is the resolution. Other popular containers are FLV, WebM and 3GPP. In Section \ref{three}, we make an analytical study of how different bit rates affect the energy consumption. In our experiments, we also use audio/videos of similar bit rates and show how they are related with the energy consumption (cf. Section~\ref{five_methodology} and~\ref{five_burst_interval}). FLV and WebM containers are exclusively used by the Flash and HTML5-based players, respectively, in Web browsers~\cite{hoque2013wowmom}. However, the iOS devices do not support FLV.

\subsubsection{Buffering and adaptive streaming} The streaming services
need to deal with bandwidth fluctuation and jitter which are prevalent
on the Internet. Therefore, the streaming clients do initial buffering
of the content at possible maximum rate. This phase is also known as Fast Start. After this phase, different approaches are used to send content to the streaming clients over
TCP. Some services send content at the encoding rate to serve more clients. Services like YouTube throttle the sending rate a little bit higher than the encoding rate to reduce the number of playback pause events due to bandwidth fluctuation. So-called Fast Caching downloads the whole content as fast as possible, which can eventually be very resource wasteful when the whole content is not consumed by the user. \citeN{guole} proposed to use a buffer adaptive mechanism, which downloads content to refill the playback buffer only when it touches a low level threshold. This mechanism creates an ON-OFF traffic pattern. \citeN{hoque2013wowmom} identified that video players in Android devices use such a method. 

The methods described above are used by video streaming services, such as YouTube, and Dailymotion, which cannot switch the video stream bit rate, i.e. quality, on the fly according to the available bandwidth. On the other hand Vimeo, Netflix and a number of IPTV services use rate-adaptive streaming in order to avoid playback pause events which annoy clients and reduce user engagement. These services apply different adaptive protocols, such as Apple's HTTP Live Streaming (HLS), Microsoft's Smooth Streaming (MSS), etc. In Section~\ref{four_three}, we design algorithms to deal with bandwidth fluctuation and to provide energy savings for both CBR and rate-adaptive streaming.

\subsection{Delivering Multimedia Content in Bursts using TCP}
\label{two_burst_tcp}
\begin{figure}[tp]
\begin{minipage}[t]{0.48\linewidth}
 \centering
   \includegraphics[width=0.8\linewidth, height=0.5\linewidth]{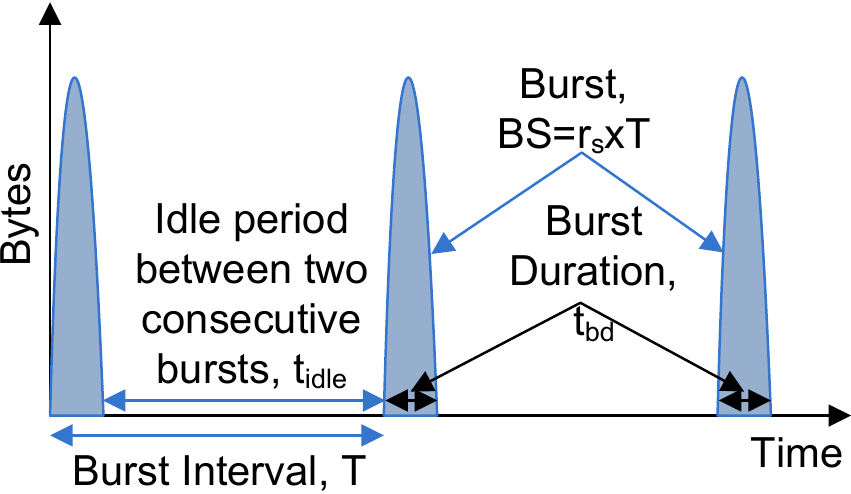}
\caption{Bursty traffic pattern and properties.}
\label{fig:burst_info}
\end{minipage}
  \hspace{0.02cm}
    \begin{minipage}[t]{0.48\linewidth}
\centering
   \includegraphics[width=0.7\linewidth, height=0.5\linewidth]{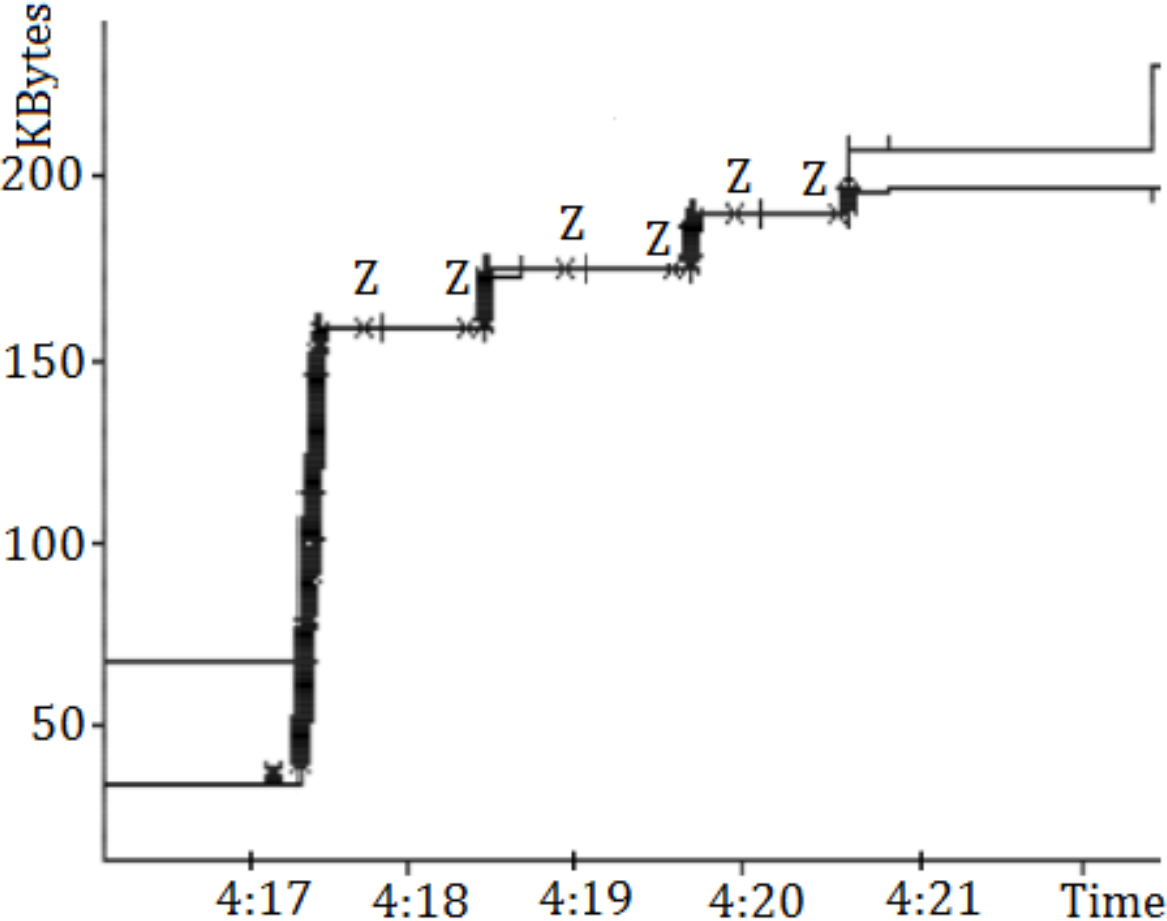}
   \caption{Burst split in constant bit rate streaming.}
    \label{fig:burst_cbr}
    \end{minipage}
    \vspace{-2mm}
\end{figure}

In Section~\ref{sub_character}, we have described that a large fraction of the content has the bit rate of a few hundred kilobits per second. On the other hand, smartphone users enjoy 2 Mbps mobile network speed on average~\cite{cisco}. Therefore, there is typically more bandwidth available on the path between the client and server than the encoding rate of the content. The WNI will be less utilized if the same amount of content can be received in bursts at the maximum possible rate instead of continuously receiving at the encoding rate. Figure~\ref{fig:burst_info} shows that if the encoding rate is $r_{s}$ and $BS$ amount of data is collected during a period of duration $T$ and then sent to the client, the client's WNI is idle most of the time and can reside in a low power state until the next burst is received. Thus, receiving the multimedia stream in the form of periodic bursts is attractive and causes the tail energy to be suffered by the client only once per each burst as opposed to being present with each packet when receiving a constant bit rate streaming. Streaming clients that use a rate-adaptive protocol also retrieve content in a similar bursty manner with a periodic interval of 2-40 seconds.

There is a caveat. Multimedia players maintain a fixed size playback buffer. The size depends on the implementation of the player and can be restricted by the operating system of the smartphone. If the size of a burst is larger than the available space in the playback and TCP receive buffer together, then TCP flow control becomes active at the client. Since the player decodes content at the encoding rate, TCP flow control and this player behavior together ensure that the excess bytes are received separately at a lower average rate. ~\citeN{huang2012} also observed similar behavior for rate adaptive streaming. From their traffic traces, we have identified that switching to an upper quality ($>$1500 kbps) triggers TCP flow control at the client when streaming from Hulu. Figure~\ref{fig:burst_cbr} demonstrates how TCP flow control at the client controls the receiving of an over-sized burst when the burst interval is of 10 s for CBR. `Z' in this figure indicates zero window advertisement from the client. The activation of TCP flow control leads to increased energy consumption, as we will show in the next section.

\begin{table}[tp]
\tbl{Description of the parameters used in the modeling and describing EStreamer. For Wi-Fi, $P_{tail}$ is the power
  draw in an idle state, i.e. radio on but no rx/tx.  For HSPA, it is the combination of the power draw in CELL\_DCH and CELL\_FACH states
  depending on the timer values and $t_{idle}$. In case of LTE (with DRX), it is the power draw from the DRX inactivity timer.} 
  {\scriptsize
      \begin{tabular}{|p{52mm}c|p{50mm}l|}
        \hline
        {Description} & {Parameter}&{Description} &{Parameter}\\
        \hline
        Avg stream encoding rate & $r_s$&Fast Start Streaming Period & $t_{fs}$\\\hline
        TCP bulk transfer capacity to client & $r_{btc}$ & Data Transferred during $t_{fs}$ & $L$\\\hline
	    Data Transferred during $t_{fs}$ & $L=r_{btc}\cdot t_{fs}$& Current burst interval & $T$\\\hline
	    Power when receiving data at rate $r$ & $P_{rx}(r)$ &Burst Duration &$t_{bd}$ \\\hline
        Power increase at rate $r$ & $\Delta P_{rx}(r)=P_{rx}(r)-P_{s}$& Energy-optimal burst interval & $T_{opt}$\\\hline
        Inactivity timer values & $T_1, T_2$ & Burst interval that avoids starvation & $T_{max}$\\\hline
        Power draw in tail energy states &$P_1, P_2$ &Idle time between two consecutive bursts & $t_{idle}$ \\\hline
        
        Avg power while spending & $P_{tail}$ & Current lower bound while autotuning $T$ & $T_{min}$\\\cline{3-4}
        tail energy(radio on but no rx/tx)& & Burst size & $BS =T \times r_{s}$\\\hline
        Tail energy & $E_{tail}(T)$&  Optimal Burst Size & $BS_{OPT}$\\\cline{3-4}
        
        
        &&Available buffer space at the client & $B$\\\hline
        
       \end{tabular}} 
    \label{tab:parameters}
\end{table}

\section{Power Modeling and Analysis of Bursty Streaming}
\label{three}
Based on the bursty streaming behavior and energy consumption
characteristics described in the previous section, we now develop
power consumption models for delivering streaming content in
bursts. We use the models to study the power consumption of different
burst sizes in different scenarios. We identify the optimal burst size
and quantify the potential energy savings and losses using the optimal
and a non-optimal size, respectively. In the previous section, we
explained that the limited buffer causing activation of TCP
flow control can take place during both constant quality and adaptive
streaming. Therefore, the models we develop are also applicable for
both kinds of streaming.

\subsection{Parameters}

The parameters used in the following equations are described in Table
\ref{tab:parameters}. We consider the power consumption to be fixed
when actively receiving data at a given rate~\cite{xiao}, and we
assume that the average stream encoding rate ($r_s$) is always lower
than the TCP bulk transfer capacity ($r_{btc}$), i.e. there is some spare
bandwidth to exploit. Following the
earlier discussions, we consider the limited buffer size of the client
($B$) and look at two corresponding cases: i) the total buffer size is
sufficiently large to fit an entire burst and ii) the buffer gets full
when receiving a burst and TCP flow control is activated. Power consumption for 
executing the application and baseline idle power of the WNIs (i.e., sleep power) are excluded from the models as
they merely add a constant overhead.

\subsection{The Receiver buffer is sufficiently large}
\label{three_three}

First, we consider the case when the fixed-size burst can be entirely
absorbed by a streaming client, i.e. $r_sT \leq B$. We obtain the
average power draw for a given $T$ according to Eq \eqref{eq:1_1},
which comprises the burst download power and the power drawn by the
tail, and its derivative with respect to $T$ according to Eq
\eqref{eq:1_2}.

{\small
\begin{align}
\label{eq:1_1}
P(T) &=\frac{Tr_s\Delta P_{rx}(r_{btc})}{Tr_{btc}} + \frac{E_{tail}(T)}{T}\\
\frac{dP}{dT} &=\frac{1}{T}\frac{dE_{tail}(T)}{dT} - \frac{E_{tail}(T)}{T^2}
\label{eq:1_2}
\end{align}}

To compute the derivative of $E_{tail}(T)$, we need to consider the
impact of different inactivity timers on the tail energy. This energy
is not always fixed but depends on $T$ when the idle time in between
receiving two consecutive bursts ($t_{idle}(T)$) is smaller than the
sum of the inactivity timers, which can happen especially with 3G. In
\eqref{eq:2_first}-\eqref{eq:2_last}, we compute $E_{tail}(T)$ and its
derivate in three different cases, because we have a maximum of two
timers (the case of HSPA), and apply them to \eqref{eq:1_2} (we skip the
straightforward manipulation steps). Wi-Fi and LTE cases
only have one inactivity timer and, hence, the resulting models can be
applied to them trivially by simplifying the equations so that
T2=0.

{\small
\begin{align}
\label{eq:2_first}
0<t_{idle}(T)<T_1:& \;\; E_{tail}(T) = P_1t_{idle}(T) = P_1T(1-\frac{r_s}{r_{btc}}) \implies \frac{dE_{tail}(T)}{dT} = P_1(1-\frac{r_s}{r_{btc}}) \implies \frac{dP}{dT} = 0\\
T_1<t_{idle}(T)<T_1+T_2:& \;\; E_{tail}(T) = P_1T_1 - P_2T_1 + P_2T(1-\frac{r_s}{r_{btc}}) \implies \frac{dE_{tail}(T)}{dT} = P_2(1-\frac{r_s}{r_{btc}})\\
                & \;\; \implies \frac{dP}{dT} = \frac{T_1}{T_2}(P_2-P_1) < 0\textrm{, since }P_2<P_1\\
T_1+T_2<t_{idle}(T):& \;\; E_{tail}(T) = P_1T_1 + P_2T_2  \implies \frac{dE_{tail}(T)}{dT} = 0 \implies \frac{dP}{dT} = - \frac{E_{tail}(T)}{T^2} < 0
\label{eq:2_last}
\end{align}
}

The above result shows that the power draw either stays the same or
decreases when $T$ is increased as long as player buffer and TCP buffer together can hold the
whole burst. Thus, the larger the value of $T$, the less energy is
consumed by the client device.

\subsection{Burst Size Exceeds TCP Receive Buffer Size}
\label{three_four}

Now, we consider the other case where a burst is larger than the available buffer space at the client,
i.e. $r_s\times T>B$. In this case, the portion of the burst that
exceeds the buffer size can be transmitted to the receiver at an
average rate of $r_s$ because that is the rate at which the
application consumes data from the buffer. Note that tail energy is no
longer dependent on $T$ and is consumed over a fixed-length interval
since the time it takes to transmit the whole burst grows at the same
rate at which $T$ is increased (transmission rate equal to encoding
rate). The reader can easily check this for each of the three
different cases of tail energy as we did in
\eqref{eq:2_first}-\eqref{eq:2_last}. Thus, we treat $E_{tail}$ as a
constant. We obtain Eqs \eqref{eq:4},\eqref{eq:42} for the power
consumption and its derivative. They now comprise three parts: the
power drawn during download of the portion of the whole burst that
fits the receiver's buffer at TCP bulk transfer rate, the power draw
during download of the leftover data at stream encoding rate, and
power drawn by the tail.

{\small
\begin{align}
\label{eq:4}
P(T) &=\frac{B\Delta P_{rx}(r_{btc})}{Tr_{btc}} + \frac{(Tr_s-B)}{Tr_s}\Delta P_{rx}(r_s) + \frac{E_{tail}}{T}\\
\label{eq:42}
\frac{dP}{dT} &=\frac{B}{T^2}(\frac{\Delta P_{rx}(r_s)}{r_s}-\frac{\Delta P_{rx}(r_{btc})}{r_{btc}}) -\frac{E_{tail}}{T^2}
\end{align}
}

The amount of time that the WNI consumes tail energy
cannot be longer than the leftover idle time after the reception of
the whole burst and before the beginning of the next burst. We express
this bound in \eqref{eq:5}. {\small$\bar P_{tail}$}
is the average power draw while consuming tail energy and when substituting it into \eqref{eq:4} and
setting {\small $\frac{dP}{dT} \geq 0$} we obtain the
inequality in \eqref{eq:6}.

{\small
\begin{align}
\label{eq:5}
E_{tail} &\leq \bar P_{tail}(T-\frac{B}{r_{btc}}-\frac{(Tr_s-B)}{r_s}) = \bar P_{tail}(\frac{B}{r_s}-\frac{B}{r_{btc}})\\
\label{eq:6}
\frac{r_{btc}}{r_s}&\geq \frac{P_{rx}(r_{btc})-\bar P_{tail}}{P_{rx}(r_{s})-\bar P_{tail}}
\end{align}
}

We now use a model written out in \eqref{eq:7} for the power consumption while receiving
content, which linearly depends on the data rate. This is in general a safe
assumption for WNI's~\cite{xiao}. Because obviously the tail power never exceeds the
power drawn while receiving at any rate, the constant $a$ must be
greater than or equal to one. When we substitute \eqref{eq:7} into \eqref{eq:6},
after some straightforward manipulation, we obtain \eqref{eq:8}.

{\small
\begin{gather}
\label{eq:7}
P_{rx}(r) = (a + kr)P_{tail}\textrm{, subject to: }  a\geq 1\\
\label{eq:8}
(a-1)(\frac{r_{btc}}{r_s}-1) \geq 0
\end{gather}
}

The inequality in \eqref{eq:8} is always true. Note that we did not even assume anything about the slope
of the linear function ($k$). Hence, we have established that when the burst size
exceeds the receiver buffer space, the power consumption is a
non-decreasing function of $T$. Therefore, the lowest energy consumption
can be achieved when the burst size exactly matches the available buffer space
at the receiver.

\begin{figure}[tp]
  \begin{center}
    \subfigure[500kbps stream over Wi-Fi.]{\label{fig:buffer_burst_power-wifi-500kbps}\includegraphics[width=0.38\linewidth, height=0.32\linewidth]{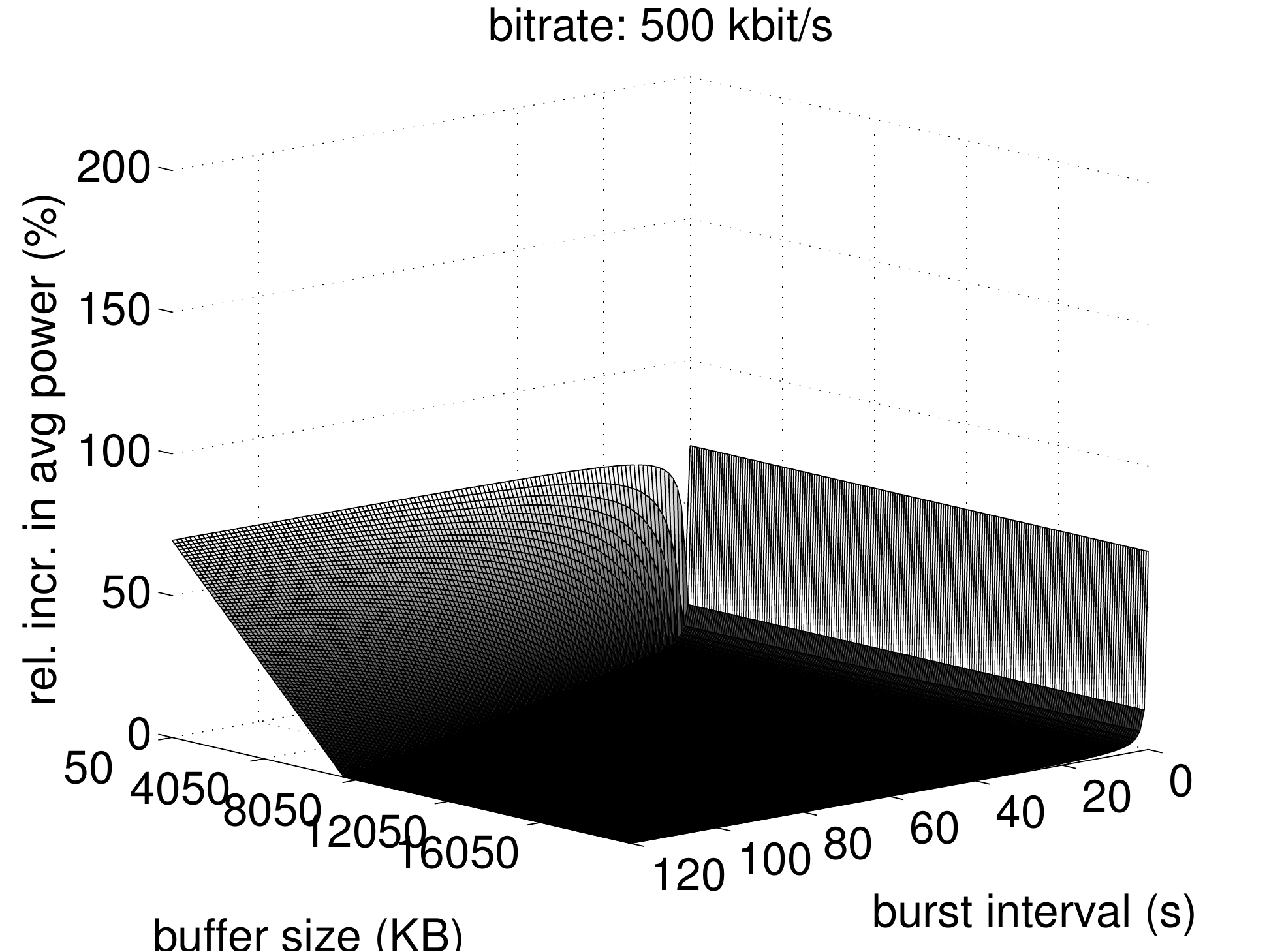}}
    \subfigure[500kbps stream over LTE.]{\label{fig:buffer_burst_power-lte-500kbps}\includegraphics[width=0.38\linewidth, height=0.32\linewidth]{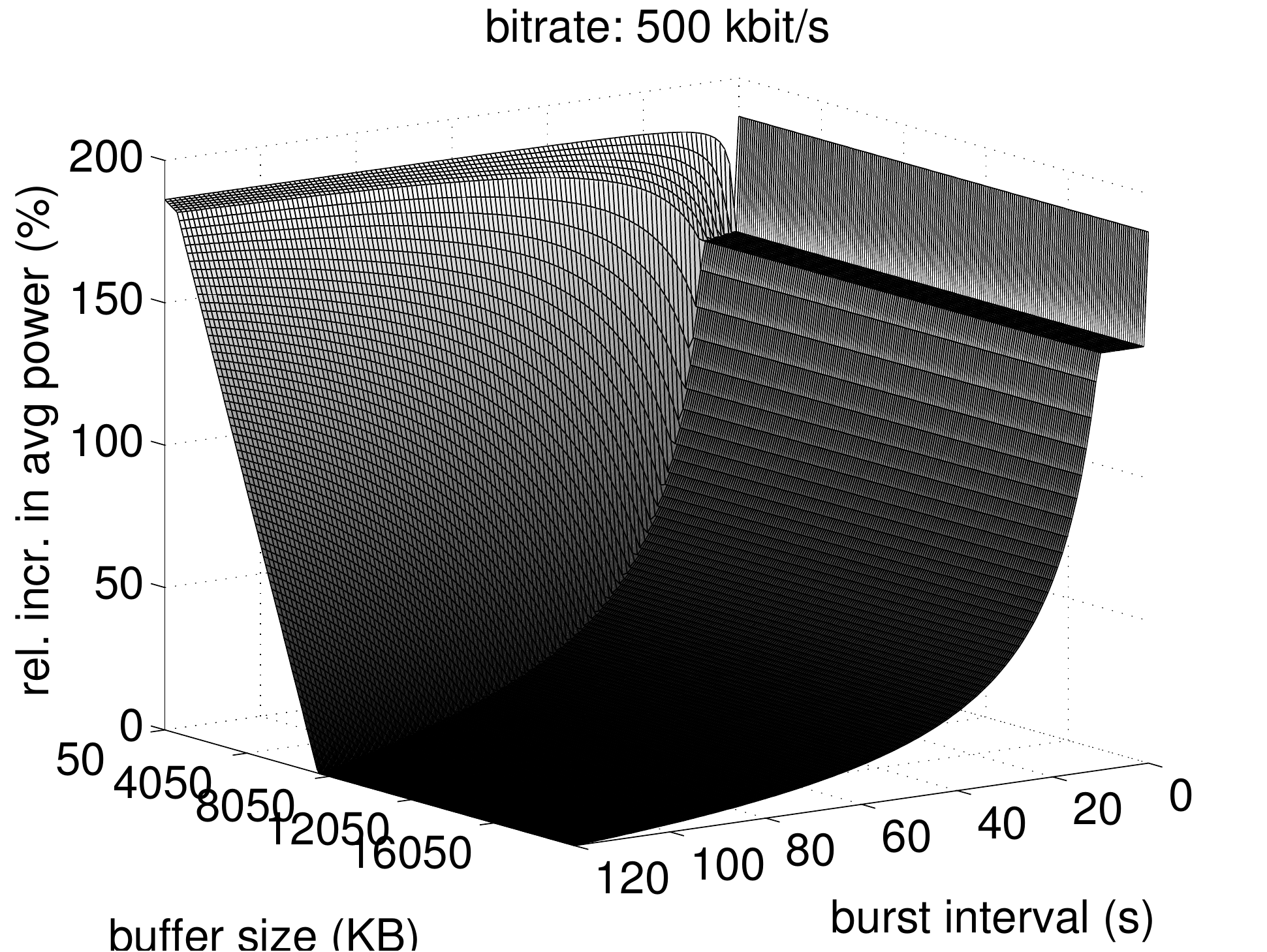}}\\
    \subfigure[10MB buffer space using Wi-Fi.]{\label{fig:bitrate_burst_power-wifi-10MB}\includegraphics[width=0.39\linewidth, height=0.32\linewidth]{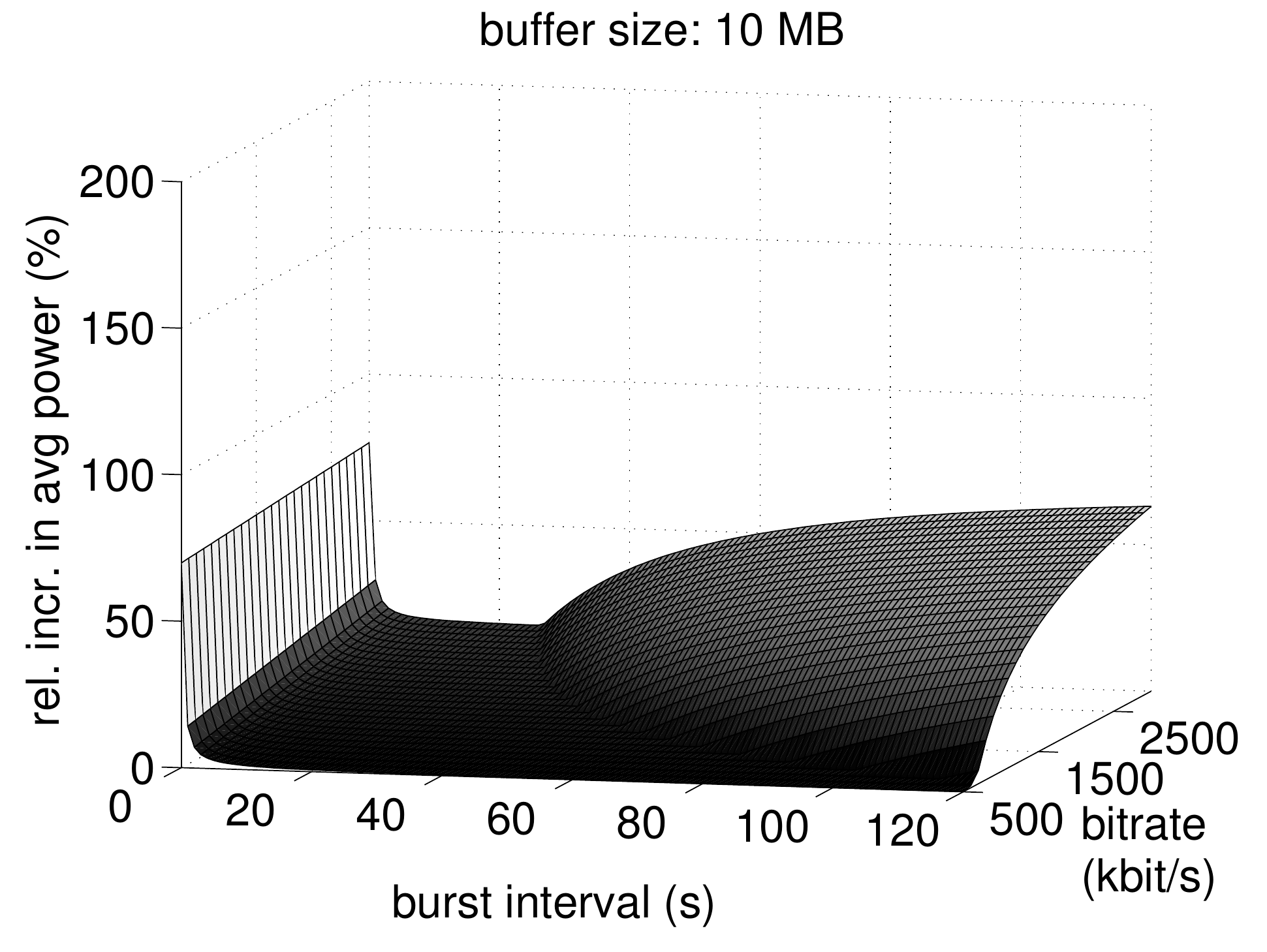}}
    \subfigure[10MB buffer space using LTE.]{\label{fig:bitrate_burst_power-lte-10MB}\includegraphics[width=0.39\linewidth, height=0.32\linewidth]{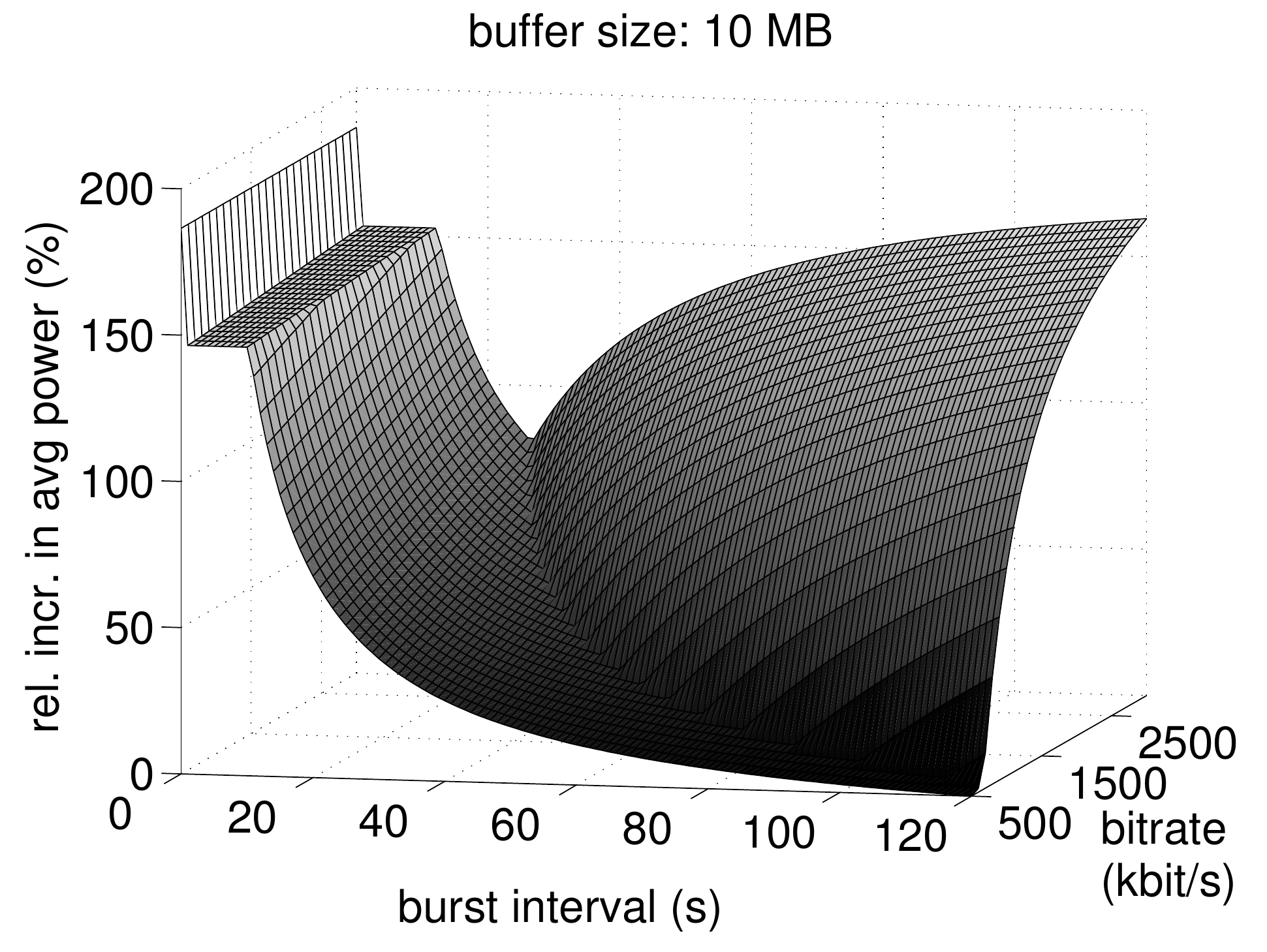}}

    \caption{Average power consumption with different burst intervals and amount of buffer space.}
    \label{fig:burst_vs_buffer}
  \end{center}
  \vspace{-7mm}
\end{figure}

\subsection{Burst Size vs. Buffer Space Analysis}
\label{sec:three_analysis}

To quantify the impact of using optimal or non-optimal burst size on
energy consumption, we study the energy consumption in a few cases
using \eqref{eq:1_2} and \eqref{eq:4}. To capture the effect of
different amounts of tail energy, we compare LTE without DRX and
Wi-Fi. HSPA results are rather similar to LTE results so we omit them
for brevity. Parameters were set according to measured values from
experiments with HTC Velocity LTE: $\Delta P_{rx}(r_{btc})$ was set
to 760 mW and 1520 mW, and $r_{btc}$ to 20 Mbit/s and 16 Mbit/s,
respectively for Wi-Fi and LTE. The tail energy ($E_{tail}$) was
computed using typical values for the inactivity timers (Wi-Fi's PSM
timer 0.2 s and LTE's inactivity timer 10 s) and $P_{tail}$ was set to
435 mW for Wi-Fi and to 1216 mW for LTE. In the case of Wi-Fi, we set
$\Delta P_{rx}(r_s)$ to a value between the idle power ($P_{tail}$)
and $\Delta P_{rx}(r_{btc})$ assuming a linear relationship with the data
rate. For LTE, we set $\Delta P_{rx}(r_s)$ to the same value as
$\Delta P_{rx}(r_{btc})$.

Figures~\ref{fig:buffer_burst_power-wifi-500kbps} and
\ref{fig:buffer_burst_power-lte-500kbps} plot the average power
consumption as a function of both burst interval ($T$) and available
buffer space ($B$). We make two main
observations. First, the average power consumption decreases much more
sharply with Wi-Fi than with LTE when the burst interval (and size) is
increased. The reason is the difference in the length of inactivity
timers: it takes a much longer burst cycle with LTE to amortize the
tail energy associated with each burst than with Wi-Fi. Note that the
10 s inactivity timer is visible as a plateau in the LTE surface plot
when the burst cycle is shorter than that. Second, and
more importantly, while setting the burst cycle to a few tens of
seconds may yield good energy savings immediately, setting the burst
size just a bit too large leads to a significant increase in power
consumption. This penalty is more striking with LTE where the power
draw does not scale with the data rate and downloading content at the
encoding rate is very energy inefficient. Hence, it is not a good idea
to just blindly choose a ``large enough'' burst size. Instead, a
smarter mechanism is required. We propose such a mechanism in the next
section.

In Figures~\ref{fig:bitrate_burst_power-wifi-10MB} and
\ref{fig:bitrate_burst_power-lte-10MB}, the client buffer size was
fixed to 10 MB while bit rate and burst size vary. We observe that the
lower the stream encoding rate, the larger the energy saving
potential. This result stems from the fact that the lower the stream
encoding rate, the more spare bandwidth is available to leverage by
the bursts. In other words, while two streams with different encoding
rates consume the same amount of energy when received at constant bit
rate (WNI continuously powered on), when they are shaped into fixed
size bursts, the one with a lower encoding rate leads to smaller size
bursts which take less time and, consequently, energy to receive than
the larger bursts of the stream with higher encoding rate. This
property is also much more pronounced with LTE compared to Wi-Fi.

\section{ES\lowercase{treamer}}
\label{four}

In this section, we propose a multimedia delivery system called EStreamer that can be integrated with an audio/video streaming service. It works with both constant quality and rate-adaptive streaming. EStreamer turns an input multimedia stream into bursts and delivers those to the client. During a streaming session, it determines the optimal burst size for the client and uses that when constructing the bursts. It is a cross-layer system and consists of two components: (1) Traffic Profiler which works at the transport layer and (2) Traffic Shaper which works at the application layer.

\vspace{2mm}
\noindent\textbf{Traffic Profiler (Profiler):} The job of the Traffic Profiler is to capture TCP acknowledgement packets coming from the streaming clients and log the arrival time of the ACKs. From the ACK packets, it collects the sequence numbers of the ACKs and advertised receive window size. The log of the ACKs arrival times are used to calculate the duration of a burst ($t_{bd}$) and the receive window size is used to estimate the total buffer, $B$, status at the streaming client. Window size zero, i.e. ZWA, indicates that the client's total playback buffer $B$ is full. The sequence number is used to determine whether a streaming client received all the packets of a burst.

\vspace{2mm}
\noindent\textbf{Traffic Shaper (Shaper):} The first task of the Shaper is to find the encoding rate of the content, $r_{s}$, by parsing the stream header. The next task is to keep track of the amount of data sent to the client during the Fast Start to calculate the maximum burst cycle, $T_{max}$. It is the maximum duration that the player can play without distorted playback unless more content is received. If the encoding rate is $r_{s}$ bytes per second and $L$ bytes are transferred to the client during $t_{fs}$, then the client has $T_{max}=\frac{L}{r_{s}}$ (assuming that $r_{s}$ is constant) duration of playback content. Finally, it decides a burst cycle, $T$, according to the player buffer status received from the Profiler. In the meantime, the Shaper continues buffering the incoming traffic and sends a burst of size $T\times r_{s}$ to the client when $T$ expires. The Shaper can also determine the end-to-end bandwidth for a burst, when it receives the corresponding burst duration from the Profiler.

\subsection{Finding the Optimal Burst Interval $T_{opt}$}
\label{four_two}

Since encoding rate, $r_{s}$, is constant for a given stream quality, the burst size only depends on $T$. The Shaper seeks an optimal burst interval, ($T_{opt}$), with the help of the Profiler. In Section~\ref{two}, we showed that the client TCP sends ZWAs to the sender to halt data transmission when the clients TCP receive buffer becomes full. The Profiler uses ZWAs to find the optimal burst size, $BS_{OPT}$, and reports to the Shaper. Then, the Shaper finds $T{opt}$, which is calculated as $T_{opt}=\frac{BS_{OPT}}{r_{s}}$. In this section, we discuss the techniques that EStreamer uses to find the optimal burst size at different phases of a streaming session.

\vspace{2mm}
\noindent\textbf{Finding the optimal burst size during Fast Start:} EStreamer begins a streaming session with Fast Start and it is possible that the Profiler would find that the client's total buffer is filled even before the Fast Start is over. This event is more likely with very high bit rate streams. In this case, $BS_{OPT}=SentBytes$. Here, $SentBytes$ is the amount of data delivered to the client till the first zero window advertisement received by the Profiler. 

\vspace{2mm}
\noindent\textbf{Finding the optimal burst size after Fast Start:} A straightforward strategy for finding such a burst size after Fast Start would be to start traffic shaping with a small value of $T$ and then to increase it gradually. The shortcoming of this approach is that it may take a very long time to find the optimal one. For example, if $T_{opt}=15 $ s then it would take 120 s to reach, whereas the on-demand videos are of a few minutes in length (see Section~\ref{sub_character}). Therefore, EStreamer uses binary search to speed up. The mechanism is presented in Figure~\ref{fig:es_cbr} and~\ref{fig:es_vbr} as flow charts for two different streaming systems.

The basic mechanism works as follows. 
The Shaper calculates $T_{max}$ during $t_{fs}$ as described earlier. Then, it begins traffic shaping. Initially, it selects $T=T_{max}/2$ seconds and starts buffering incoming traffic from the streaming server over this period. The Profiler checks for the ZWAs in the traffic profile and reports to the Shaper. If there are no ZWAs, then $T$ is increased. Finally, there will be one of the following two outcomes; (I) EStreamer reaches $T_{max}$ without experiencing any ZWA and finds $BS_{OPT}$, (II) EStreamer observes ZWAs and $BS_{OPT}=SentBytes$ for a T which is less than $T_{max}$.

\subsection{Dealing with Bandwidth Fluctuation}
\label{four_three}

\begin{figure}[tp]
\begin{minipage}[t]{0.5\linewidth}
\includegraphics[height=0.9\linewidth, width=1\linewidth]{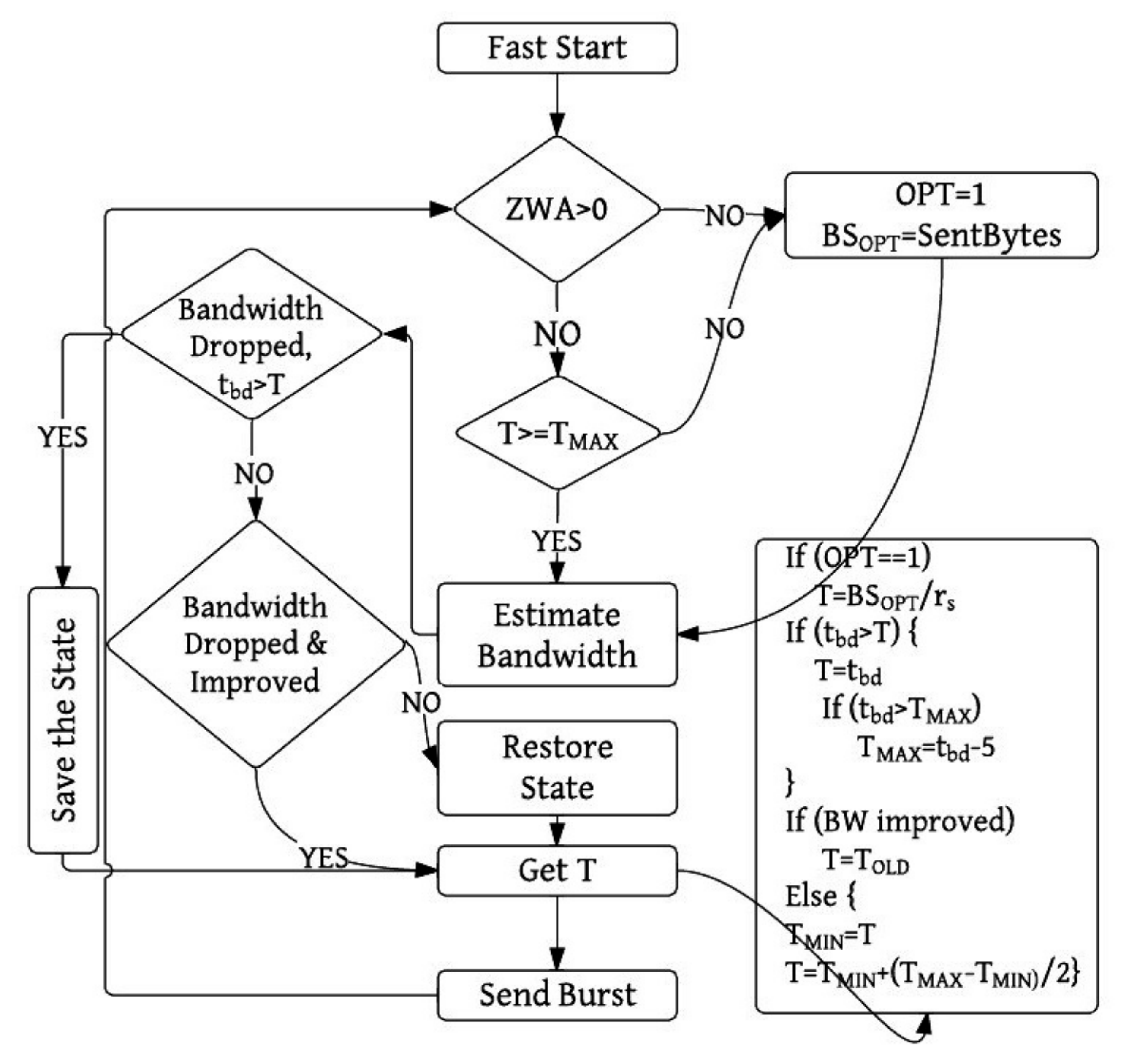}
\caption{EStreamer's operation to find the optimal burst size for CBR streaming.}
\label{fig:es_cbr}
\end{minipage}
\hspace{0.1mm}
\begin{minipage}[t]{0.5\linewidth}
\includegraphics[height=0.9\linewidth, width=1\linewidth]{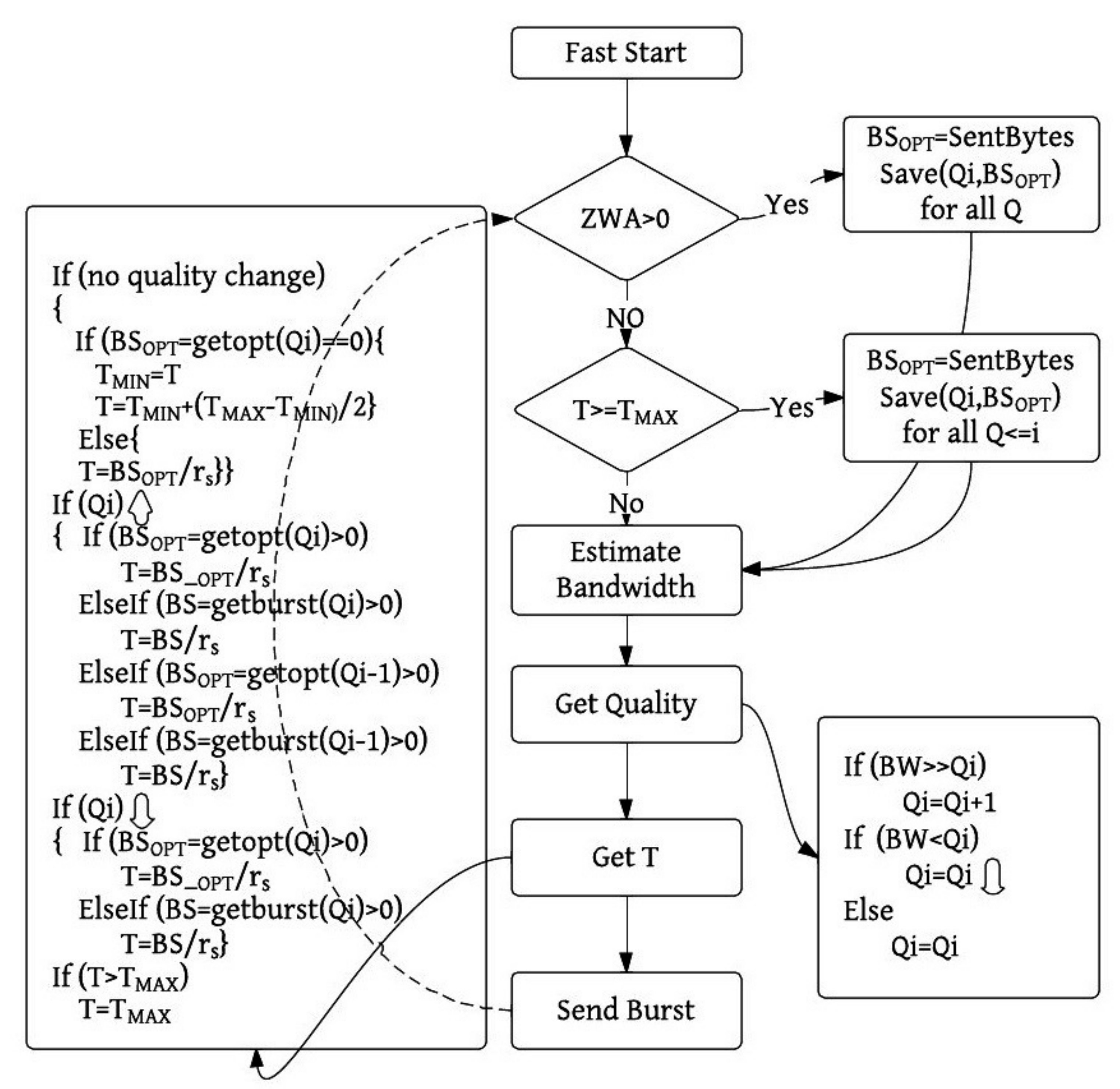}
\caption{EStreamer's operation to find the burst interval for optimal power consumption for rate-adaptive streaming.}
\label{fig:es_vbr}
\end{minipage}
\vspace{-2mm}
\end{figure}

The playback at the streaming clients can be interrupted when using CBR streaming applications even without EStreamer if the bandwidth drops below the encoding rate. This situation can happen when traffic shaping is in one of the following three states: (i) $BS_{OPT}$ has not been found yet, (ii) $BS_{OPT}$ has been found and limited by the client's buffer space, or (iii) $BS_{OPT}$ has been found and limited by $T_{max}$ (see Figure~\ref{fig:es_cbr}). EStreamer preserves one of these occurred states by setting $T_{old}=T$. After that, it continuously sends content to the client while measuring the bandwidth in order to detect when the situation improves. It also continuously updates $T_{max}$ based on the sent content and the encoding rate as long as the low bandwidth situation continues. It restores the old state by setting $T=T_{old}$, when the bandwidth situation improves. After that it begins traffic shaping again to find the optimal burst size.

Rate-adaptive streaming is the appropriate solution to deal with bandwidth fluctuations. Such mechanisms are DASH~\cite{Stockhammer}, HLS, and MSS. Although these mechanisms employ their own rate adaptation algorithms at the clients, we implement the algorithm at the EStreamer. One of the main reasons is that the earlier version of EStreamer already had bandwidth detection and traffic shaping mechanisms~\cite{hoque2013nossdav}. We only added the quality switching functionality based on the detected bandwidth. 

Figure~\ref{fig:es_vbr} shows how EStreamer applies traffic shaping and rate adaptation together. The stream has been encoded into $n$ different qualities which each have a different encoding rate $r_s^i, i\in {1,2,3...n}$. At the beginning, EStreamer serves a streaming client with the maximum quality, the encoding rate of which is less than or equal to the half of 2 Mbps (i.e. $r_s^i\leq (1/2 \times 2Mbps)$). This initial choice is based on the finding by~\citeN{cisco} that smartphones enjoy on average 2 Mbps mobile network speed. In addition, TCP provides satisfactory
performance when the end-to-end bandwidth is twice the encoding rate,
considering the effect of packet loss, congestion, round trip time,
and playback rate at the client~\cite{wang08tomccap}. If the client does not send any ZWA during the Fast Start phase, the content sent in that phase marks the largest possible burst size that can be used. Hence, EStreamer sets this burst size as the $BS_{OPT}$ for the current quality and computes $T_{max}=T_{opt}=\frac{BS_{OPT}}{r_{s}}$. It also sets this burst interval ($T_{opt}$) to be the optimal for any other lower quality stream with inferior encoding rate. The reason is that when switching to a lower quality stream, the receive buffer cannot become a bottleneck because the size of the content for the same length (in seconds) of burst would be smaller. However, the same burst interval may not be optimal for a higher quality stream because it is possible that the buffer space at the client becomes a limiting factor when the encoding rate increases since the amount of bytes also increases. Therefore, when switching to a higher quality stream, EStreamer uses the previously found $BS_{OPT}$ to determine an equivalent $T$ and continues traffic shaping using binary search again. If ZWAs are observed, then the EStreamer sets the optimal burst size for all qualities. The simple quality-switching algorithm is shown in Figure~\ref{fig:es_vbr}. A switch happens only if the determined average bandwidth permits. EStreamer allows upgrading the quality only when the current bandwidth is at least twice the encoding rate of the higher quality. The reason for this restriction is discussed earlier.

Different policies for quality switching can be plugged into the
adaptive part of EStreamer. We want to emphasize that our simple
quality switching policy may be far from optimal in some cases and that our goal is
to demonstrate how adaptive streaming can be supported in
EStreamer.

\subsection{Implementation}
For using with popular constant bit rate streaming services, such as YouTube, we chose to deploy EStreamer in an HTTP proxy server and placed the proxy in the cloud. The reason is that we had no possibility of deploying EStreamer on the corresponding servers. The smartphones' proxy settings were configured to use EStreamer. EStreamer does not shape traffic during the Fast Start phase and forwards 20-45 s playback data depending on the service. 

In order to experiment rate-adaptive streaming, we integrated EStreamer with a streaming server and developed a separate client. The communication between the client and EStreamer works based on HTTP over TCP. We describe the HTTP request/response headers in appendix \ref{invalid}. At the beginning, the client makes the initial GET request. EStreamer selects the initial quality and responds by sending the stream initialization header of the corresponding quality. It also mentions the duration of the video, bit rate, time range, and other parameters in the response header (\textsf{X-Stream-Info: duration=597;bitrate=700000;seconds=0-59;height=480; width=853}). Next, the client makes the request for actual content in time range (\textsf{Range: seconds=0-}) to which EStreamer responds by sending the next chunk. Note that the client specifies only the start of the range. It is the job of EStreamer to determine an appropriate burst size and whether the stream quality should be switched or not.  After that, whenever the client's playback buffer goes down to only 5 seconds, it sends a new request specifying the next missing content as the beginning of the range (e.g. \textsf{Range: seconds=60-}). If EStreamer decides to switch stream quality, it specifies the new bit rate in the response header (\textsf{X-Stream-Info: duration=597;bitrate=2000000; height=720; width=1280; seconds=100-139;}). However, response mismatch can occur when there are ZWAs from the client as the EStreamer might send less content than the time range specified in the header. This is because EStreamer aborts sending the remaining content of the determined burst size. In this case, the client again requests the remaining content and EStreamer responds with the response code 204 and correction in the response header as shown in appendix~\ref{invalid3}.

\begin{table}[tp]
  \begin{center}
	    \tbl{The streaming services, the encoding rate of the streams, frame rates, container, and their total playback durations.}
    {\scriptsize
      \begin{tabular}{|c|c|c|c|c|}
        \hline
        \textbf{Streaming Service}& \textbf{Bit rate (kbps)}& \textbf{Frame Rate} & \textbf{Container} & \textbf{Duration}\\\hline
        
        ShoutCast(audio)(CBR) & 128 &--& MP3 & 4-10 \\\hline
        YouTube Browser (CBR) & 280,328,2000 &25& FLV,MP4& 4:47,9:57 \\\hline
        YouTube App (CBR) & 458,2000 &25,24& MP4 & 4:47,9:57 \\\hline
        Dailymotion App (CBR) & 452 &25& MP4 & 5:58 \\\hline
        	Service X  & 700,1200,1500,2000,2500,3000 &	96&	MP4	& 9:57\\\hline
    \end{tabular}}
    \label{tab:streaming_pro}
  \end{center}
\end{table}

\subsection{Discussion}

We briefly discuss here the impact of different
characteristics of various multimedia streams on the behavior of
EStreamer. It can support different encoding rates, codecs, and
containers as long as the
streaming content is carried by HTTP over TCP. In Hulu traces, we have seen that TCP flow control can also take place if the client cannot accommodate a chunk when using HTTP rate-adaptive streaming. Therefore, EStreamer logic can be integrated with DASH-like clients as well. However, such a solution would have information about the playback buffer and TCP receive buffer.  Consequently, the heuristic can be bypassed.

EStreamer counts in bytes when searching for the
optimal burst size and then computes the burst interval. Therefore, the mechanism is robust also when variable
bit rate (VBR) streaming is used. Since the traffic shaper
uses the average bit rate to compute the corresponding burst interval
during which the target sized burst is generated, there may be cases
where the burst size is not exactly of the right size if the encoding
rate deviates significantly from the average during that
interval. Hence, the resulting burst may be slightly bigger or smaller
than the optimal size in some cases. In some streaming systems, the server can send audio and video as two separate streams. In such a case, EStreamer must synchronize the shaping of
the two streams so that the burst transmissions coincide in order to
maximize the idle time in between the burst transmissions. The mechanism
to detect the optimal burst size remains the same and will be done
separately for the two streams.

\section{Performance Evaluation}
\label{five}

We used three popular constant bit rate multimedia streaming services, Internet Radio, YouTube, and Dailymotion, in four different smartphones via EStreamer. 
Table~\ref{tab:streaming_pro} shows the streaming services and the
properties of the multimedia contents used in the smartphones in our experiments. A
128-kbps Internet radio channel was played in all the smartphones and two different YouTube videos were viewed. One video was encoded at 280 and 320 kbps rates with FLV container. We also streamed a 458 kbps version of the same video of MP4 container. The rate of the other video was 2000 kbps and the duration was ten minutes. A 452 kbps video was streamed from Dailymotion using the native application. For rate-adaptive streaming (Service -X in Table~\ref{tab:streaming_pro}), we used the DASH data set of ``big buck bunny'' from~\citeN{Lederer:2012} . 

\subsection{Methodology}
\label{five_methodology}

\begin{table}[tp]
  \vspace{-1mm}
  \begin{center}
	    \tbl{Test cases with different network parameter
      configurations.}
    {\scriptsize
      \begin{tabular}{|c|c|c|}
        \hline
        \textbf{RAN} & \textbf{Test cases} & \textbf{Network configuration}\\
        \hline
        \multirow{3}{*}{HSPA} & default(def) & T1=8s, T2=3s, T3=29mins, CELL\_PCH on\\
        \cline{2-3}
        & aggressive(aggr) & T1=6s, T2=2s, T3=29mins, CELL\_PCH on\\
           \cline{2-3}
        & no PCH(noPCH) & T1=8s, T2=10s, CELL\_PCH off\\
        \hline
        \multirow{3}{*}{LTE} & default without DRX & RRC$_{idle}$=10s, DRX off\\
        \cline{2-3}
        & default with DRX & RRC$_{idle}$=10s, DRX$_{idle}$=750ms, DRX$_{c}$=640ms, DRX$_{on}$=20 ms\\
        \cline{2-3}
        & default with DRX, long idle & RRC$_{idle}$=20s, DRX$_{idle}$=750ms, DRX$_{c}$=640ms, DRX$_{on}$=20ms\\
        \cline{2-3}
        \hline
    \end{tabular}}
    \label{tab:nw_test_cases}
  \end{center}
  \vspace{-4mm}
\end{table}

We did two sets of measurement. In the first set, we used Wi-Fi,
a commercial HSPA 3G network and the LTE 4G test network of Nokia Solutions and 
Networks. The smartphones were connected to the Internet via a DLink
DIR-300 wireless AP supporting 802.11 b/g. The bandwidth of the Wi-Fi
network was 54 Mbps. In the case of HSPA and LTE, the configurations used were the ``def" configuration and default configuration with DRX, respectively, in Table~\ref{tab:nw_test_cases}. The results obtained from this measurement setup are presented in Section~\ref{five_burst_interval} and~\ref{background}. We also describe the results obtained during bandwidth fluctuation scenarios via HSPA and LTE in Section~\ref{five_bandwidth_fluc}.

The second set of measurements was done in the HSPA and LTE test networks of NSN with different timer settings. Table \ref{tab:nw_test_cases} summarizes the network configurations for the different test cases. For HSPA measurements, we used three different network configurations. Default configuration refers to configuration according to the vendor recommended parameters. Aggressive configuration refers to shorter values of T1 and T2. In the third configuration, the T2 value was set to a longer value when the CELL\_PCH state was disabled. LTE experiments were done with cDRX disabled or enabled and with short or long inactivity timer, RRC$_{idle}$,  values. The purpose of these tests is twofold: (i) study the effect of timer settings on the energy consumption of smartphones (Section~\ref{five_impact_cellular}) and (ii) study the effect of our energy-aware traffic shaping on radio network signaling (Section~\ref{five_radio_signaling}).

We used Monsoon power monitor and Nokia Energy Profiler to measure the power consumption of the smartphones. During the audio streaming sessions the displays were turned off. For video streaming sessions, the displays were on. Therefore, the video measurement results include power consumption for decoding, display and wireless networking. We present the energy savings result in the presence of EStreamer by comparing with the power consumption of mobile devices when EStreamer was absent.  

We present our measurement results and analysis in three steps. First,
we evaluate the performance of EStreamer in determining the optimal burst
cycle and providing energy savings. We also discuss how EStreamer deals with bandwidth variations in the network. Second, we illustrate the effect of various cellular network configurations on energy savings. Finally, we consider the impact of traffic shaping on radio network signaling.

\begin{table}[tp]
  \tbl{Maximum power savings at $T_{opt}$ for different mobile phones using EStreamer, while playing CBR audio and video streaming applications over Wi-Fi, HSPA and LTE. YouTube videos were played using both browser (``Bro'') and the native application. The LTE measurement results are only from the HTC Velocity phone.}
   {\scriptsize
    \begin{tabular}{|p{40mm}|p{23mm}|c|c|c|}
      \hline
      Applications&  HTC Nexus One & Nokia N900 & Nokia E-71 &HTC Velocity\\
	  Access Network & (Android-2.3.6) & (Maemo) & (Symbian V 3.0)&(Android 2.3.7 \\
	  & Sav\%--kbps--$T_{opt}$ & Sav\%--kbps--$T_{opt}$ & Sav\%--kbps--$T_{opt}$&Sav\%--kbps--$T_{opt}$\\
     \hline\hline
      Internet Radio/Wi-Fi &23\%--128--14 s($\star$)& 62\%--128--14 s($\star$)& 65\%--128--6 s($\diamond$)& -- \\\hline
      Internet Radio/HSPA/LTE&38\%--128--14 s($\star$)& 24\%--128--14 s($\star$) & 2\%--128--4 s($\diamond$)& 60\%--128--18 s($\star$)\\\hline
      YouTube Bro/Wi-Fi& 14\%--328--39 s ($\star$)& 20\%--328 --39 s ($\star$)& 18\%--280--4 s($\diamond$) &\\\hline
      YouTube Bro/HSPA/LTE& 14\%--328--39 s ($\star$)& 14\%--328--39 s ($\star$)& 4\%--280--3 s($\diamond$)&50\%-2000-31 s($\circ$)\\\hline
      YouTube App/ Wi-Fi&13\%--458--39 s($\star$)&--& --&--\\\hline
      YouTube App/HSPA/LTE& 27\%--458--39 s($\star$)&--& --&54\%--2000--39 s($\star$)\\\hline
      Dailymotion/Wi-Fi& 15\%--452--33 s($\diamond$)&--& --&--\\\hline
      Dailymotion/HSPA/LTE& 30\%--452--33 s($\diamond$)&--& --&55\%--452--33 s($\diamond$)\\\hline
      \end{tabular}}
    \label{tab:power_savings}
\end{table}

\subsection{Burst Size Tuning and Energy Consumption for Constant
  Quality Streaming}
\label{five_burst_interval}

In this section, we demonstrate how EStreamer finds the optimal burst
size when streaming constant quality content. We observed three
different cases: In the first one, EStreamer detects ZWAs during the
Fast Start phase of streaming a 2 Mbps video to the browser-based
player in the HTC Velocity phone. As a result, the traffic shaper determines
$T_{opt}=31$ s and always sends a burst after every 31 s till the end
of the video (see Table~\ref{tab:power_savings}). In the second case,
EStreamer determines $T_{opt}$ using the binary search ending up with
a value that is less than $T_{max}$, which happens when streaming the
Dailymotion video. These scenarios are indicated with a diamond symbol in the Table. Third, EStreamer finds $T_{opt}$ at $T_{max}$ when streaming the 2 Mbps video to the native YouTube player in HTC Velocity and these results are indicated with a star symbol.

One observation from the results is that streaming the same bit rate
audio via HSPA to N900 saves 24\% energy, whereas the Nexus One saves 38\%
energy. The reason is that N900 can spend only
(14-$t_{bd}$-T1-T2)$\approx$2 seconds in CELL\_PCH state, whereas the
Nexus One uses legacy Fast Dormancy with an inactivity timer of 6.5 s
and spends (14-$t_{bd}$-6.5)$\approx$7 seconds in the IDLE state. We
discuss the effect of FD and different timer settings in cellular
networks in Section~\ref{five_impact_cellular}. It is also evident
that the potential energy savings are larger with audio than with
video streaming, which stems from the fact that the data transfer
energy consumption makes a larger component in the total consumption
in audio streaming compared to video streaming where display and
decoding take a bigger share. Finally, the results show that the
largest energy savings can be achieved when streaming via Wi-Fi (65\%),
which is followed by LTE (55\%) and HSPA (38\%). These savings improve
the battery lifetime of Nokia E-71, HTC Velocity, and Nexus One by 3x,
2x and 1.5x, respectively (see Table~\ref{tab:battery_impro}).

\subsection{Impact of Background Traffic}
\label{background}

\begin{table}[tp]
  \begin{center}
	    \tbl{Improvement in battery life time in hours using EStreamer.}
    {\scriptsize
      \begin{tabular}{|c|p{22mm}|p{10mm}|p{12mm}|p{8mm}|p{15mm}|p{15mm}|p{15mm}|}
        \hline
        \textbf{Device}& \textbf{Savings\%}& \textbf{Battery\break (mAh)}&\textbf{(mA)\break no EStr.} & \textbf{(mA)\break EStr.} &\multicolumn{3}{|c|}{\textbf{Battery life (h)}}\\
        \cline{6-8}
        &&&&&\textbf{no EStr.}&\textbf{EStr.}&\textbf{bg traffic}\\\hline
        E-71 (Wi-Fi)& 65\%--128--6 s($\diamond$)&1500 & 270 & 85 &5.1&14.7&- \\\hline
        Velocity (LTE)& 50\%-2000-31 s($\circ$)&1620 & 622 & 311 & 2.6 & 5.2&-\\\hline        
        Nexus (HSPA)& 38\%--128--14 s($\star$)&1400& 405& 251 & 3.5 & 5.5&-\\\hline
        Nexus (HSPA)& 14\%--328--39 s($\star$)&1400& 480& 407 & 2.92 & 3.42&3.27\\\hline
        N900 (Wi-Fi)&  62\%--128--14 s($\star$) &1320& 270& 103 &4.9&12.8&-\\\hline
        N900 (HSPA)&  20\%--328--39 s($\star$) &1320& 540& 432 &2.4&3.0&2.4\\\hline        
    \end{tabular}}
    \label{tab:battery_impro}
  \end{center}
  \vspace{-4mm}
\end{table}

\begin{figure}[t]
\centering
\includegraphics[height=0.28\linewidth,width=0.65\linewidth]{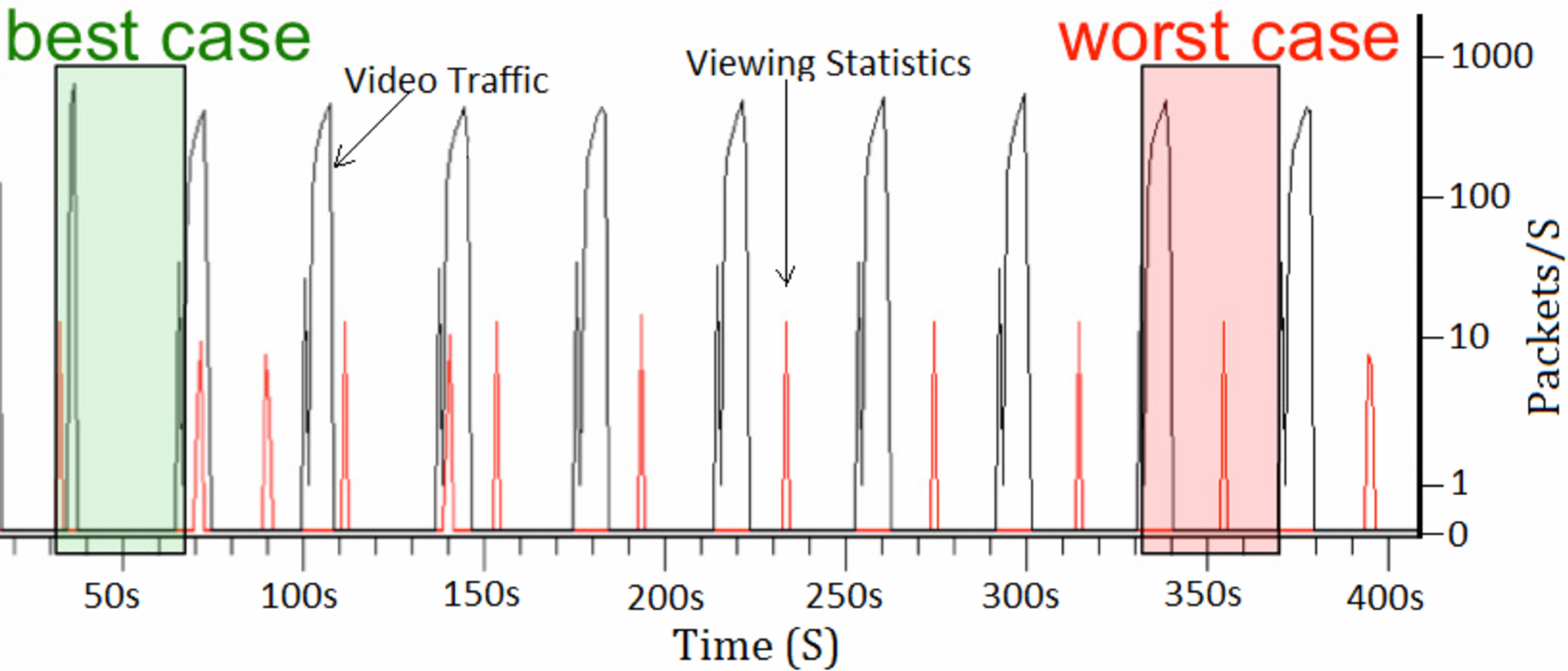}
\caption{Isolating best and worst case bursts for background traffic impact estimation. The Y axis is in logarithm scale.}
\label{fig:best_worst}
\vspace{-2mm}
\end{figure}

Surprisingly, Table \ref{tab:power_savings} indicates that streaming a low quality video (328 kbps) from YouTube to browser (``Bro'') in Nexus One consumes more energy than streaming a higher quality video (458 kbps) to the native application via HSPA in the same device. Figure~\ref{fig:best_worst} illustrates what happens during YouTube streaming to the browser in Nexus One via EStreamer, as an example. The browser-based player establishes several TCP connections to YouTube servers. EStreamer shapes only the traffic of the connection that carries the video content (black spikes). The other connections are used to report the user's viewing progress (red spikes).

Other applications installed in a smartphone can also generate such periodic background traffic. The prime examples are the email clients and free applications with embedded advertisement code of some advertisement platform, such as AdMob. We observed that the email client in an Android phone fetches email every five minutes. The interval for retrieving an advertisement is within the range of 12-120 s~\cite{Prochkova:2012}. The packets of these applications can interleave with the video content bursts in an unfortunate manner and the impact is an increase in overall energy consumption. We measured an increase of 27\% and 6\% power consumption in N900 and Nexus One, respectively, in the worst case (in presence of a background connection which reports the user's viewing history). In Nexus One, we observed similar increase when the email client retrieves email from the server. Table~\ref{tab:battery_impro} shows that background traffic can nullify the achievement from traffic shaping if the worst-case scenario appears during each burst interval.

\subsection{Bandwidth Fluctuation: Constant Quality \& Rate-Adaptive Streaming}
\label{five_bandwidth_fluc}

\begin{figure}[tp]
\begin{minipage}[t]{0.325\linewidth}
\centering
\includegraphics[width=1.0\linewidth, height=0.7\linewidth]{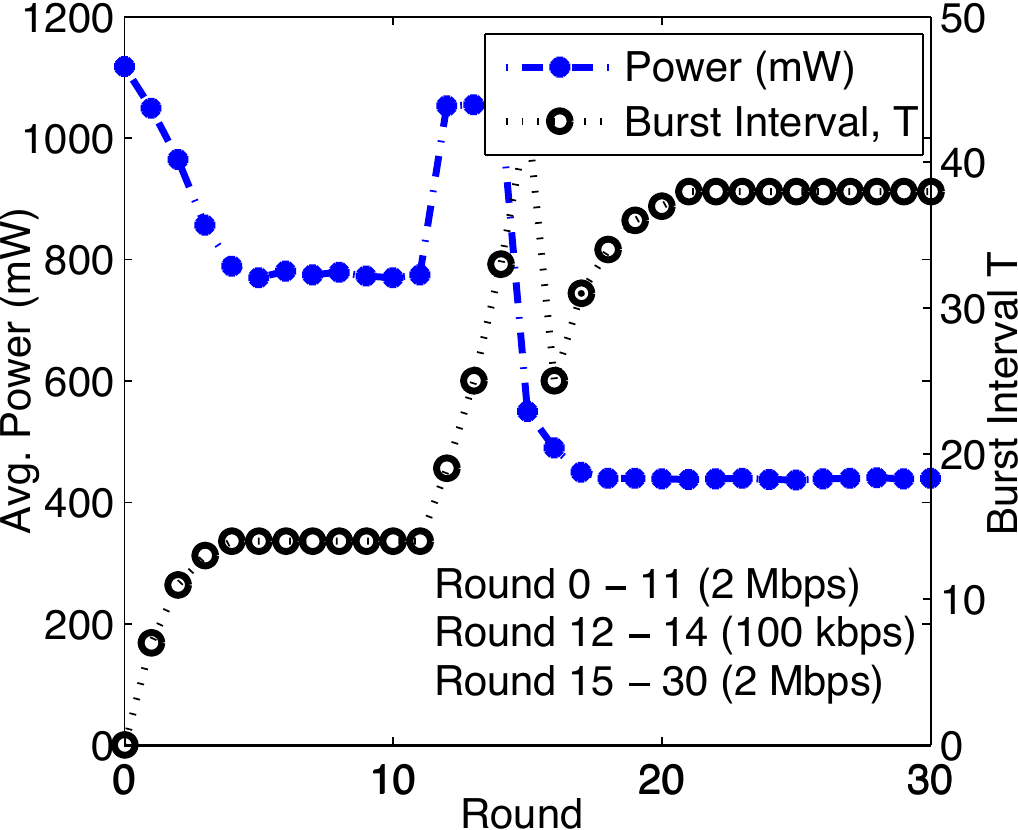}
\caption{Power consumption of N900 during bandwidth variation and streaming a 128 kbps audio stream via HSPA.}
\label{fig:bandvar1}
\end{minipage}
\hspace{1mm}
\begin{minipage}[t]{0.325\linewidth}
\centering
\includegraphics[width=1.0\linewidth, height=0.7\linewidth]{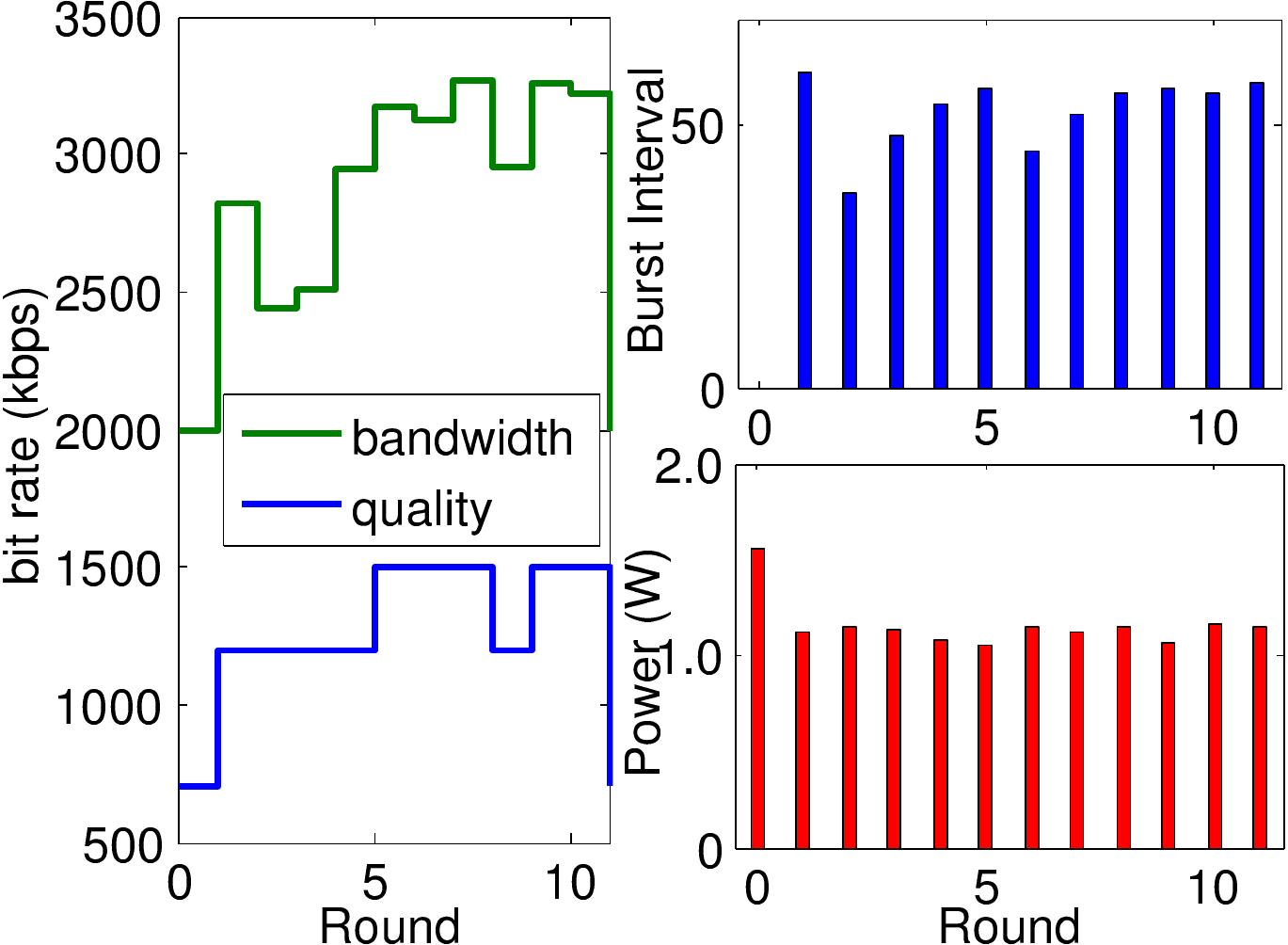}
\caption{Impact of bandwidth variation on quality, burst interval, and power of the HTC Velocity when streaming via HSPA.}
\label{fig:bandvarG}
\end{minipage}
\hspace{1mm}
\begin{minipage}[t]{0.325\linewidth}
\centering
\includegraphics[width=1.0\linewidth, height=0.7\linewidth]{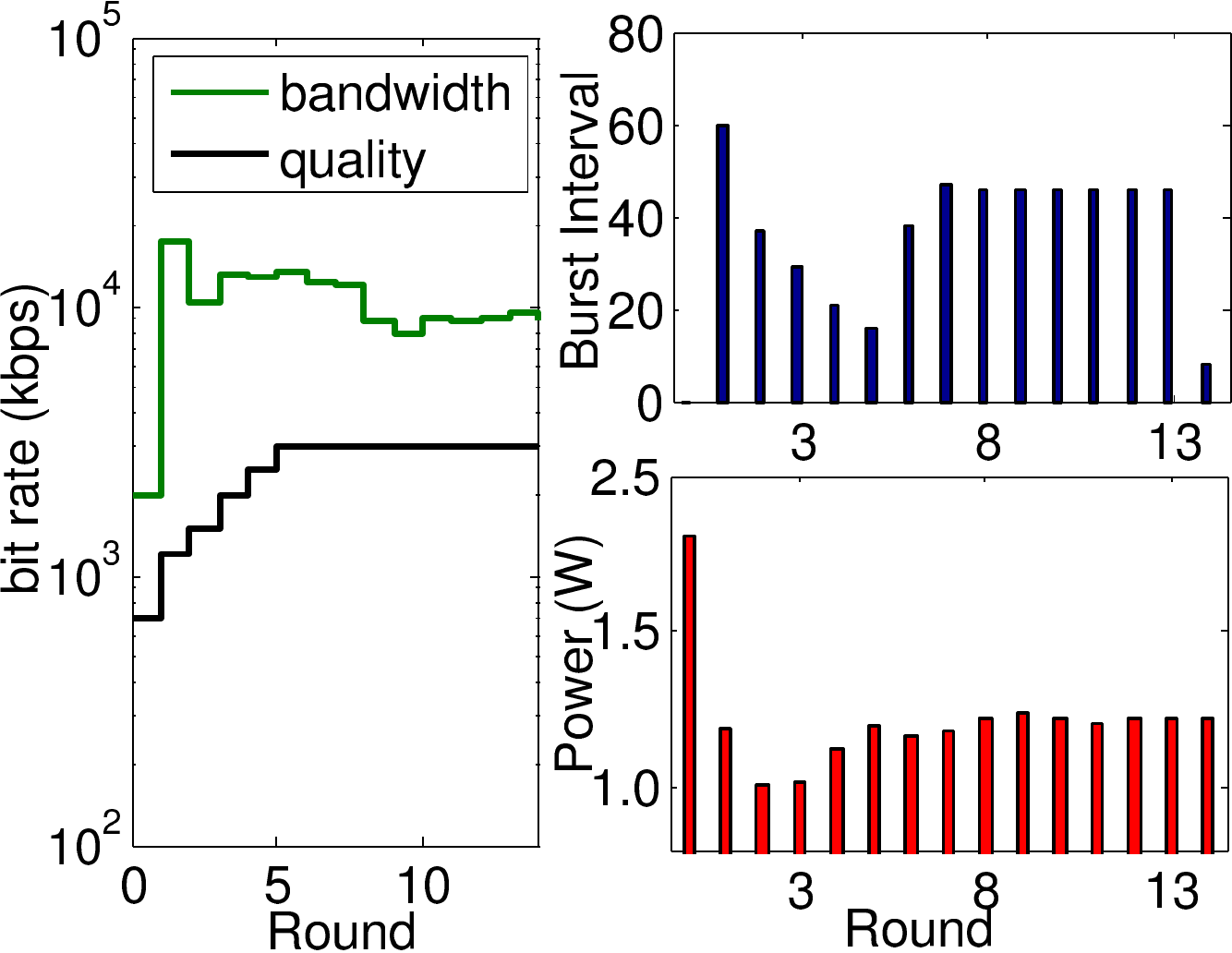}
\caption{Impact of bandwidth variation on quality, burst interval and power of the HTC Velocity when streaming via LTE.}
\label{fig:bandvarL}
\end{minipage}
\vspace{-2mm}
\end{figure}

In this section, we describe how EStreamer deals with bandwidth variability. For streaming audio of constant quality, we used~\citeN{wondernew} at the EStreamer hosting machine to limit the downloading bandwidth of the streaming client. Figure~\ref{fig:bandvar1} shows a streaming session in N900 via HSPA. It shows that power consumption increases as the bandwidth decreases after the 12th round. It also illustrates when bandwidth falls below the encoding rate, there is playback interruption. In this case, EStreamer saves this state by setting $T_{old}=14$ s. After a few rounds, $T_{max}$ increases to 43 s. From the 15th round bandwidth improves and EStreamer reintroduces the old state, $T=T_{old}=14$ s, and continues to shape traffic again.

We also experimented with rate-adaptive streaming when streaming via HSPA by moving around the campus during the streaming sessions. We found that EStreamer fluctuated among the three lowest qualities. This is because of the limited downlink capacity of the subscription. Since it was not possible to measure power consumption while roaming, we present a similar scenario in Figure~\ref{fig:bandvarG}. The stair graph in the figure shows that EStreamer switches to the maximum 1500 kbps stream at the 5th round, and transitions to the 1200 kbps at the 7th round. At the 9th round, EStreamer switches back to the 1500 kbps. The stair graph in Figure~\ref{fig:bandvarL} shows how EStreamer changes to a higher quality when there is enough bandwidth using LTE in log scale. At the 5th round it switches to the 3000 kbps stream. Therefore, from the upper bar chart we can see that initially the burst interval decreases as the quality of video increases. EStreamer increases the burst interval by applying binary search and finds the optimal burst interval for 3000 kbps video at $T=46$ s, as the TCP flow control reacts. It sets the corresponding burst size as $BS_{OPT}$ for all other qualities. Then, it continues traffic shaping with the same burst interval and quality till the end of the video. The lower bar chart shows that power consumption is less when streaming lower quality videos. The key messages from these measurement results are the following. Power consumption increases when bandwidth decreases. In an extreme situation, the bandwidth may decline below the lowest available encoding rate and a user may experience distorted playback. However, EStreamer can react accordingly for both constant bit rate and rate-adaptive streaming for improving the battery life of a smartphone.

\subsection{Impact of Cellular Network Configuration}
\label{five_impact_cellular}

Table~\ref{tab:energy_config} presents results obtained with different network configurations. We first note that shorter timers do not generally help to save energy in the absence of EStreamer. If EStreamer is used, slightly more energy can be saved with more aggressive timer values than the default configuration. Energy savings for audio streaming sessions are higher than the video streaming scenarios. The reason is the quicker transition to lower power state after receiving a burst. However, this observation does not hold for video streaming to Nexus One as the device uses legacy Fast Dormancy. From traffic and power traces we identified that it uses an inactivity timer value of approximately 6.5 s. This timeout is shorter than T1 values in default and no CELL\_PCH scenarios but it is just a bit longer than the T1 value in the aggressive setup. As a consequence, Nexus One activates the legacy FD in most of the cases. Background traffic also contributes to the energy consumption increase, which we already explained in Section~\ref{background}.

\begin{table}[tp]
\vspace{-2mm}
          \tbl{Energy savings of smartphones when streaming video and audio with different HSPA network configurations.} 
  {\scriptsize
      \begin{tabular}{|p{34mm}|p{17mm}|p{16mm}|p{17mm}|p{18mm}|}
        \hline
        & \multicolumn{2}{|c|}{\textbf{YouTube}} & \multicolumn{2}{|c|}{\textbf{Audio}} \\\cline{2-5}
        \textbf{Network Configuration} & \textbf{Nokia N900} & \textbf{Nexus One}&\textbf{Nokia N900} & \textbf{Nexus One} \\\hline  
        default & 14\% & 16\%&24\%&38\% \\\hline
        no PCH & 11\% & 20\% &18\%&38\% \\\hline
		aggressive &19\%& 12\%&46\%&43\% \\\hline        
       \end{tabular}}
     \label{tab:energy_config}
\end{table}

\begin{figure}[tp]
\begin{minipage}[b]{0.32\linewidth}
\centering
\includegraphics[scale=0.4]{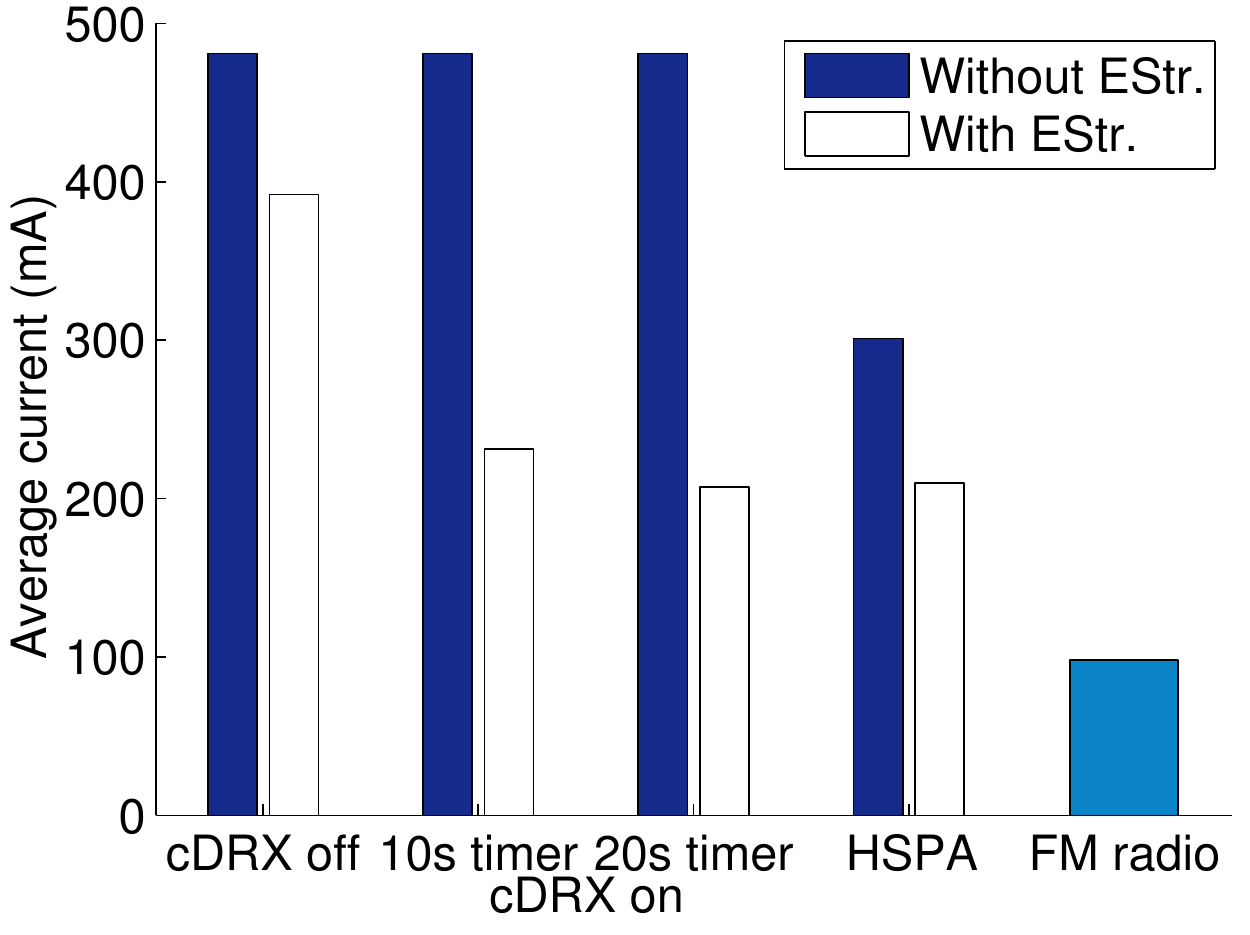}
\caption{Energy consumption while streaming audio via HSPA and LTE.}
\label{fig:audio_LTE}
\end{minipage}
\hspace{0.01cm}
\begin{minipage}[b]{0.33\linewidth}
\centering
\includegraphics[width=1.0\textwidth,height=0.7\textwidth]{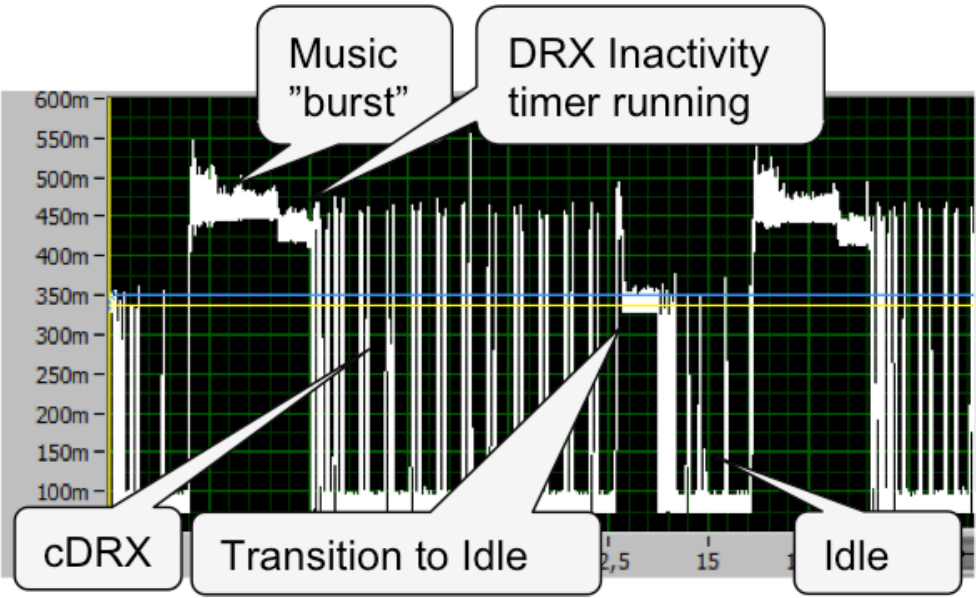}
\caption{DRX w/ 10s inactivity timer. X-axis is time (s) and Y-axis current (mA).}
\label{fig:audio_LTE_drx}
\end{minipage}
\hspace{0.01cm}
\begin{minipage}[b]{0.33\linewidth}
\centering
\includegraphics[width=1.0\textwidth,height=0.7\textwidth]{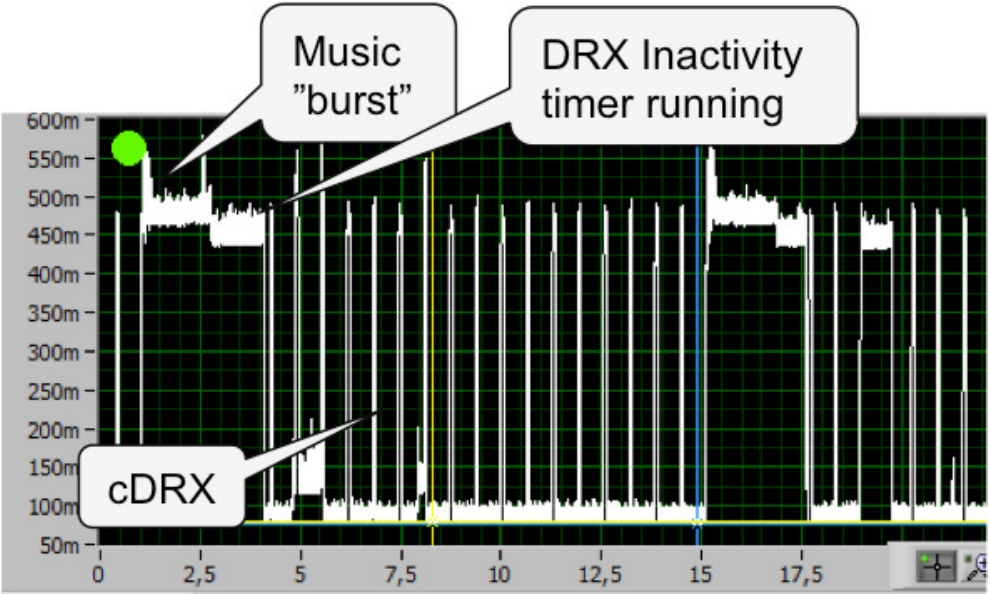}
\caption{DRX w/ 20s inactivity timer. X-axis is time (s) and Y-axis current (mA).}
\label{fig:audio_LTE_drx20}
\end{minipage}
\vspace{-5mm}
\end{figure}

Energy savings with different LTE parameter configurations are displayed in Figure~\ref{fig:audio_LTE}. The figure shows that larger savings can be obtained by increasing the inactivity timer value from 10 s to 20 s, which seems counter-intuitive at first. When RRC$_{idle}= 10$ s, the LTE protocol transitions from RRC\_CONNECTED to RRC\_IDLE in between receiving the bursts because the timer is shorter than the burst interval. We can also see that without EStreamer streaming over LTE requires more energy than over HSPA and streaming FM radio. In the case of HSPA, the energy savings is relatively lower than the LTE scenarios with EStreamer. The reason is the long inactivity timers. Figure~\ref{fig:audio_LTE_drx} shows that this transition causes a non-negligible amount of energy to be spent due to the signaling. If RRC$_{idle}$ is increased to 20 s, this transition no longer occurs and thus the average current is decreased (see Figure~\ref{fig:audio_LTE_drx20}). Even further savings should be possible when increasing the burst interval and optimizing the cDRX profile.

\subsection{Impact of Traffic Shaping on Radio Network Signaling}
\label{five_radio_signaling}

The energy savings achieved by reshaping traffic into bursts come with
a price when using cellular network access. The savings are achieved by increasing the number of
RRC state changes during the streaming session. Each state change
generates some signaling traffic. We extracted the number of state
changes for each test case from the signaling logs. Since we know
the number of signaling messages required for a specific state change,
we were able to compute the total number of signaling messages.

The signaling measurement results in the HSPA network for YouTube traffic shaping are shown on the left-hand side of Table \ref{tab:youtube_signaling}. Although in each case EStreamer causes more signaling load compared to the case without EStreamer, there is a clear difference between Nexus One and Nokia N900. The reason lies in the legacy FD that Nexus One uses. Since that mechanism terminates the RRC connection, the IDLE$\rightarrow$CELL\_DCH transition in between each video content burst requires a lot more signaling than the  CELL\_PCH$\rightarrow$CELL\_DCH transition because of the RRC reconnection. For the same reason, the signaling load is higher with Nexus One when CELL\_PCH is not available in the network. YouTube background traffic plays a negative role in signaling, as well. In the case of Nexus One, those periodic packets emerge right in the middle of the video bursts that causes extra IDLE$\rightarrow$CELL\_DCH$\rightarrow$IDLE transitions each time. However, N900 shows a surprisingly higher increase in signaling for no CELL\_PCH and the aggressive timers cases than the default settings. The reason turned out to be that N900 applied legacy FD after some bursts. The logic explaining how N900 decides when to apply that mechanism remains unclear to us.

\begin{table}[tp]
         \tbl{Increase in signaling in the HSPA network due to YouTube video and ShoutCast audio traffic shaping by EStreamer. The results are expressed as ``messages without EStreamer$\rightarrow$ messages with EStreamer $\uparrow$ factor of increase in number of messages'' per minute.} 
  {\scriptsize
      \begin{tabular}{|p{34mm}|p{17mm}|p{17mm}|p{17mm}|p{18mm}|}
        \hline
        & \multicolumn{2}{|c|}{\textbf{YouTube}} & \multicolumn{2}{|c|}{\textbf{Audio}} \\\cline{2-5}
        \textbf{Network Configuration} & \textbf{Nokia N900} & \textbf{Nexus One}&\textbf{Nokia N900} & \textbf{Nexus One}\\\hline  
        default & 7$\rightarrow$26($\uparrow$3.7x) & 11$\rightarrow$100($\uparrow$9.1x) & 3$\rightarrow$49($\uparrow$16x)&3$\rightarrow$114($\uparrow$38x) \\\hline
        no PCH & 6$\rightarrow$47($\uparrow$7.8x) & 11$\rightarrow$96($\uparrow$8.7x) &4$\rightarrow$49($\uparrow$11x)&3$\rightarrow$102($\uparrow$34x)\\\hline
		aggressive &6$\rightarrow$37($\uparrow$6.2x)& 14$\rightarrow$109($\uparrow$7.8x) &--&6$\rightarrow$74($\uparrow$12x)\\\hline   
       \end{tabular}}
    \label{tab:youtube_signaling}
\end{table}

We looked at the increase in signaling traffic for the audio streaming
case as well (right-hand side of Table~\ref{tab:youtube_signaling}). The main insight is that, overall, the signaling traffic grew more compared to the YouTube experiments as EStreamer selected shorter burst intervals in the audio streaming case because of the shorter $T_{max}$. Another notable difference to the video streaming results is that disabling the CELL\_PCH state did not increase the signaling load as much as in the YouTube experiments. The reason is that $T_{opt}$ is shorter than the inactivity timer values T1+T2 (8+10) s, which means that no transitions to the IDLE state happened, except with Nexus One. In the case of LTE, there are only two states and there can be two kinds of state transitions. When RRC$_{idle}=10$ s and cDRX is disabled the average number of state transitions is 4.3/min.  Signaling does not increase when cDRX is enabled and the inactivity timer is RRC$_{idle}=10$ s or 20 s. Thus the cDRX mechanism does not impose extra signaling on the network.

The above observations are important for network operators while configuring
network parameters and for researchers or mobile
vendors while designing an energy-aware mobile network access system
or policy in general. The key implications are the following: (i) The
signaling load in the network increases if a network does not support
the CELL\_PCH state. The reason is frequent RRC reconnection, (ii) The
usage of legacy Fast Dormancy by the smartphones reduces energy
consumption but increases the signaling load very rapidly as the
mobile device frequently closes and re-establishes the RRC
connection, and (iii) The usage of very short timers reduces energy
consumption but can increase the signaling load in the network because
of the frequent state transitions but the effect is less pronounced
compared to the two previous cases. In summary, if the network
supports CELL\_PCH, then the short timers or FD Rel-8.0 can be a
win-win situation for both parties as the number of signaling messages
can be tolerable for the network operators~\cite{nsn2011report} and
energy savings are significant for the smartphones. In case of LTE,
enabling cDRX does not increase in network signaling. We note that
similar discontinuous reception/transfer mechanisms exist also for HSPA
where they are applied in the CELL\_DCH state. The concept is called
Continuous Packet Connectivity (CPC) and it was introduced already in
the 3GPP Release 7. However, the support in phones and networks does
not seem widespread.

\section{Related Work}
\label{nine}

Traffic shaping is widely used to save communication energy for wireless multimedia streaming~\cite{hoque12survey}. ~\citeN{9.wang} proposed an adaptive streaming technique for the UDP-based multimedia streaming. The system works at the server or proxy and manipulates the burst interval depending on the packet loss and network situation experienced by the streaming client. On the other hand, our focus is HTTP over TCP-based streaming. We modeled the energy consumption of bursty TCP traffic. Then we developed EStreamer based on the models. EStreamer depends on standard TCP properties to apply traffic shaping, which greatly simplifies the implementation of energy aware streaming. It does not require either the support of any secondary protocol such as RTCP or the modification of TCP or any other protocols. Therefore, EStreamer can be easily integrated with modern TCP-based streaming services.

Other approaches identify idle periods at different phases of TCP-based applications, such as in the middle of the data transmission~\cite{lin,enhua}. Another example is choking/unchoking the TCP receive window size to make the TCP traffic bursty~\cite{yanc}. In this case, the burst interval is the duration between a choking and unchoking period. Authors in~\cite{enhua} applied this trick for multimedia streaming services such as RealNetwork, Windows Media and YouTube with a burst interval of 200 ms. However, these mechanisms cannot be applied for 3G and 4G, because these solutions are wireless access technology dependent as they force the Wi-Fi interface into sleep state and such an operation on cellular network interfaces would bar the smartphone from basic phone functioning. In contrast, EStreamer is independent of the WNI being used for streaming. Recently, \citeN{limm2012} proposed GreenTube to save communication energy for YouTube using multiple TCP connections. However, such approaches cannot be successful in reducing energy consumption when the application receives content at the server controlled lower  rate~\cite{hoque2013wowmom}.

Many papers have also studied the energy efficiency of 3G communication. \citeN{xiao08youtube} were among the first to study the energy consumption of YouTube streaming over both Wi-Fi and 3G. We go much beyond the scope of that work by considering the impact of traffic shaping and different network configurations, including LTE. Afterwards, \citeN{balasubramanian09imc3g} performed a measurement study on the energy consumption of 3G communication but did not consider streaming applications in their study. \citeN{qian11mobisys} characterized the energy efficiency of several different applications. They observed that some music streaming applications behave in an energy inefficient manner due to the CBR traffic. Earlier in \cite{qian10imc}, the same authors proposed a traffic shaping scheme for YouTube and compute estimates on potential energy savings with that scheme. However, they did not consider the consequence of TCP flow control on the energy consumption of WNIs in their study. A more recent and thorough measurement study on different mobile video streaming services and the resulting energy consumption on different mobile OSs and devices is presented in \cite{hoque2013wowmom}.

Concerning the RRC parameter configuration in cellular networks, ~\citeN{lee2004wts} first proposed to tune the inactivity timers dynamically. ~\citeN{qian10imc} suggested traffic aware inactivity timer configuration to reduce energy consumption. They also proposed to trigger Fast Dormancy based on the information provided by different applications~\cite{qian2010icnp}.~\citeN{falaki2010imc} proposed the total of T1+T2 = 4.5 s based on their observation on packet inter-arrival time in traffic traces.~\citeN{ukhanova} also suggested aggressive timer configuration for CBR video transmission from mobile devices.~\citeN{Deng} proposed to initiate FD dynamically instead of a fixed timeout value. In fact, with these proposed timer settings a mobile device would be able to save more energy in the presence of EStreamer. At the same time, these studies do not consider the consequence of the proposed configurations on the network and these are the cases where our study also can have significant implications. We considered the benefits and disadvantages of different network configurations for bursty streaming traffic, and thus the importance of our study also lies in careful configuration of the RRC parameters in the network and designing energy-aware network access.

Compared to our earlier work~\cite{hoque2013nossdav}, our notable new contributions are the following. First we demonstrate the relationship between available buffer space at the client, the received burst size and the energy consumption. After that, we illustrate the implementation of EStreamer for both CBR and rate-adaptive streaming, and then demonstrate how EStreamer quickly fine-tunes the energy-optimal configuration for a given client using a binary search approach. In~\cite{matti2013movid}, we analyzed only video streaming results, whereas in this work we include power and signaling measurement results both for audio and video streaming scenarios.

\section{Conclusions}
\label{ten}

In this article, we first modeled the energy consumption of TCP-based bursty streaming traffic. Then, based on the insights gathered through the modeling, we designed and implemented an energy efficient multimedia delivery system called EStreamer. We evaluated its performance and energy savings of smartphones. We showed that EStreamer strives for as large energy savings as possible for each client without compromising the quality of the streaming service. This energy efficiency is irrespective of the WNI being used for streaming. To carry out these tasks, it uses a heuristic derived from the models and it does so by shaping traffic for each client in an energy-optimal way. Finally, the energy consumption overall differs quite a lot between the devices, which is mostly explained by the different hardware components used in the phones. An important lesson from this study is that while a similar magnitude of savings can be achieved in different ways, from the network's perspective there is a big difference because of varying amounts of signaling load.

In section~\ref{background}, we have seen that the share of background traffic in the total energy consumption of smartphones is significant. Those applications retrieve information after some periodic intervals. Different applications maintain different frequency of periodicity. As a result, the network access becomes frequent and sporadic. We have witnessed how such background traffic can diminish the effect of multimedia traffic shaping. \citeN{Deng} proposed to accumulate background traffic from different applications and then send together. Their solution did not consider any networking application, specifically VoIP or multimedia, running in the foreground. In addition, modern mobile platforms, such as Android, do not provide API to hint at any application that some other application is using networking resources. However, such hints could be useful in scheduling background traffic with respect to the traffic of VoIP or other streaming applications in order to minimize the negative effect that we experienced.

\section*{ACKNOWLEDGEMENTS}
{
	This work was supported by the Academy of Finland:
grant number 253860 and FIGS. We also thank
Te-Yuan Huang from Stanford University, USA for sharing Hulu traces with us.
}


\renewcommand{\baselinestretch}{0.85}
\bibliographystyle{ACM-Reference-Format-Journals}
\bibliography{tmc,pub_sources,web_sources,siekkine}

\elecappendix

\section{Client and EStreamer Communication over HTTP for Rate-Adaptive Streaming}
\label{invalid}
\subsection{Start of a Streaming session}

\textit{HTTP Request}

{\small
\begin{lstlisting}
GET /BigBuckBunny HTTP/1.1
Host: www.service-x.com
Range: seconds=0-
\end{lstlisting}}

\noindent\textit{HTTP Response Stream Initialization}

{\small
\begin{lstlisting}
200 OK
Content-Type=video/mp4
Content-Length = 128000
Content-Range: bytes 0-127999/128000
X-Stream-Info: duration=597;bitrate=700000;seconds=0-;
height=480;width=853
\end{lstlisting}}

\noindent\textit{HTTP Request for Content}

{\small
\begin{lstlisting}
GET /BigBuckBunny HTTP/1.1
Host: www.service-x.com
Accept: */*
Range: seconds=0-
X-Device: ANDROID
\end{lstlisting}}

\noindent\textit{HTTP Response}

{\small
\begin{lstlisting}
200 OK
Content-Type=video/mp4
Content-Length= 5162500
Content-Range: seconds 0-59/60
X-Stream-Info: duration=597;bitrate=700000;seconds=0-59;
height=480;width=853
\end{lstlisting}}

\subsection{No Change in Quality}
\label{invalid1}

\noindent\textit{HTTP Request for Content}
{\small
\begin{lstlisting}
GET /BigBuckBunny HTTP/1.1
Host: www.service-x.com
Accept: */*
Range: seconds=59-
X-Device: ANDROID
\end{lstlisting}}

\noindent\textit{HTTP Response for Content}

{\small
\begin{lstlisting}
206 OK Partial Content
Content-Type=video/mp4
Content-Length: 3500000
Content-Range: seconds 60-99/40 
X-Stream-Info: duration=597;bitrate=700000;seconds=60-99;
height=480;width=853 
\end{lstlisting}}

\subsection{Change in Quality}
\label{invalid2}

\noindent\textit{HTTP Request:}
\begin{lstlisting}
GET /BigBuckBunny HTTP/1.1
Host: www.service-x.com
Accept: */*
Range: seconds=100-
X-Device: ANDROID


\end{lstlisting}

\noindent\textit{HTTP Response for Stream Initialization}
{\small
\begin{lstlisting}
200 OK
Content-Type=video/mp4
Content-Length= 128000
Content-Range: bytes 0-127999/128000
X-Stream-Info: duration=597;bitrate=2000000;seconds=100-;
height=720;width=1280 
\end{lstlisting}}

\noindent\textit{HTTP Request for Content}
{\small
\begin{lstlisting}
GET /BigBuckBunny HTTP/1.1
Host: www.service-x.com
Accept: */*
Range: seconds=100-
X-Device: ANDROID
\end{lstlisting}}

\noindent\textit{HTTP Response for Content}

{\small
\begin{lstlisting}
206 OK
Content-Type=video/mp4
Content-Length: 10000000 
Content-Range: seconds 100-139/40 
X-Stream-Info: duration=597;bitrate=2000000;seconds=100-139;
height=720;width=1280
\end{lstlisting}}

\subsection{Change in Content Range Due to Zero Window Advertisement}
\label{invalid3}
\noindent\textit{HTTP Request}
{\small
\begin{lstlisting}
GET /BigBuckBunny HTTP/1.1
Accept: */*
Host: www.service-x.com
Range: seconds=135-
X-Device: ANDROID
\end{lstlisting}}

\noindent\textit{HTTP Response}

{\small
\begin{lstlisting}
204 OK
X-Stream-Info: duration=597;bitrate=2000000;seconds=100-134;
height=720;width=1280 
\end{lstlisting}}

\end{document}